\documentclass[a4paper,fleqn,usenatbib]{mnras}

\usepackage{newtxtext,newtxmath}
\usepackage[T1]{fontenc}
\usepackage{ae,aecompl}

\usepackage{graphicx}	% Including figure files
\usepackage{amsmath}	% Advanced maths commands
\usepackage{amssymb}	% Extra maths symbols
\usepackage{enumitem}
\usepackage{wrapfig}
\usepackage{float}
\usepackage{subfigure}
\usepackage{alphalph}
\newcommand{\pp}{\hbox{$ \\ \\ ^{\prime\prime}$}}
\newcommand{\ppdot}{\hbox{$.\!\!^{\prime\prime}$}}
\newcommand{\Mbh}{M$_\textrm{BH}$}
\newcommand{\Lagn}{L$_\textrm{AGN}$}
\newcommand{\Msun}{M$_{\odot}$}
\newcommand{\Lsun}{L$_{\odot}$}
\newcommand{\sigmab}{$\sigma_b$}

\title[ACA observations of FIR-bright quasars]{On the multiplicity of ALMA Compact Array counterparts of far-infrared bright quasars}

\author[Hatziminaoglou et al.]{E. Hatziminaoglou$^{1}$\thanks{E-mail: ehatzimi@eso.org}
D. Farrah$^{2,3}$,
E. Humphreys$^{1}$,
A. Manrique $^{4}$,
I. P\'erez-Fournon$^{5,6}$,
\newauthor
L.~K. Pitchford$^{7}$,
E. Salvador-Sol\'e $^{4}$,
L. Wang $^{8,9}$
\\
$^{1}$ESO, Karl-Schwarzschild-Str. 2, 85748 Garching bei M\"unchen, Germany\\
$^{2}$Department of Physics and Astronomy, University of Hawaii, 2505 Correa Road, Honolulu, HI 96822, USA\\
$^{3}$Institute for Astronomy, 2680 Woodlawn Drive, University of Hawaii, Honolulu, HI 96822, USA\\
$^{4}$Institut de Ci\`encies del Cosmos, Universitat de Barcelona (UB-IEEC), E-08028 Barcelona, Spain \\
$^{5}$Instituto de Astrof\'{i}sica de Canarias, E-38205 La Laguna, Tenerife, Spain \\
$^{6}$Departamento de Astrof\'{i}sica, Universidad de La Laguna, E-38206 La Laguna, Tenerife, Spain \\
$^{7}$Department of Physics, Virginia Tech, Blacksburg, VA 24061, USA\\
$^{8}$SRON Netherlands Institute for Space Research, Landleven 12, NL-9747 AD, Groningen, The Netherlands\\
$^{9}$Kapteyn Astronomical Institute, University of Groningen, Postbus 800, 9700 AV, Groningen, The Netherlands\\
}

\date{Accepted XXX. Received YYY; in original form ZZZ}

\pubyear{2018}

\begin{document}
\label{firstpage}
\pagerange{\pageref{firstpage}--\pageref{lastpage}}
\maketitle

\begin{abstract}
We present ALMA Atacama Compact Array (ACA) 870 \micron \, continuum maps of 28 infrared-bright SDSS quasars with {\it Herschel}/SPIRE detections at redshifts 2 -- 4, the largest such sample ever observed with ALMA. 
The ACA detections are centred on the SDSS coordinates to within 1\pp \, for about 80 per cent of the sample. Larger offsets indicate that the far-infrared (FIR) emission detected by {\it Herschel} might come from a companion source. The majority of the objects ($\sim$70 per cent) have unique ACA counterparts within the SPIRE beam down to 3\pp-- 4\pp \, resolution. Only a 30 per cent of the sample show clear evidence for multiple sources with secondary counterparts contributing to the total 870 \micron \, flux within the SPIRE beam by at least 25 per cent.  
We discuss the limitations of the data based on simulated pairs of point-like sources at the  resolution of the ACA and present an extensive comparison of our findings with recent works on the multiplicities of sub-millimetre galaxies. We conclude that, despite the coarse resolution of the ACA, our data support the idea that, for a large fraction of FIR-bright quasars, the sub-mm emission comes from single sources. Our results suggest that, on average, optically bright quasars with strong FIR emission are not triggered by early-stage mergers but are instead, together with their associated star formation rates, the outcome of either late-stage mergers or secular processes. 
\end{abstract}

\begin{keywords}
quasars: general -- galaxies: starburst -- galaxies: star formation -- infrared: galaxies
\end{keywords}

%%%%%%%%%%%%%%%%%%%%%%%%%%%%%%%%%%%%%%%%%%%%%%%%%%

%%%%%%%%%%%%%%%%% BODY OF PAPER %%%%%%%%%%%%%%%%%%

\section{Introduction}\label{intro}

Intense star formation and accretion onto supermassive black holes (SMBH) residing at the centres of galaxies are known to coexist across all redshifts \citep[e.g.][]{farrah03, alexander05, hernan09, hatzimi10, pitchford16}, up to extremely high active galactic nuclei (AGN) and starburst luminosities  \citep[e.g.][]{omont01, farrah02, magdis14}. Both processes peak at a redshift around 2 and decline rapidly towards lower redshifts \citep[e.g.][]{aird15,fiore17}. 

Meanwhile, in the local Universe, the mass of SMBHs, \Mbh, has been shown to correlate with properties of the galaxies they are hosted in, such as the stellar velocity dispersion in the bulge, \sigmab, as well as the mass of the bulge \citep[e.g.][]{magorrian98,ferrarese00,tremaine02}. And while all this seem to indicate that the two processes are related to the cold gas reservoir, the connection between star formation and AGN activity in galaxies remains unclear. 
What is also unclear is whether star formation events, possibly occurring at kpc scales, and AGN activity, occurring at sub-pc scales from the SMBH, can directly affect one another, as observational studies of quenching remain inconclusive, even though indirect manifestations of feedback, such as large molecular gas outflows are now found routinely \citep[e.g.][]{spoon13, gonzalez14,tadhunter14,cicone14,cicone15,feruglio17}.

Other than a causality between accretion and star formation, the existence of the \Mbh \, -- \sigmab \, relation also implies that the brightest quasars might be triggered by major mergers \citep[see][]{hopkins09}. Major mergers are mechanisms that are efficient in both building up the bulge (and subsequently deepening the gravitational potential around the SMBH maintaining a high Eddington ratio accretion) and lowering the angular momentum of gas, allowing it to reach the SMBH faster than a dynamical time.
Indeed, simulations of merging systems often find star formation events reaching or exceeding 1000 \Msun yr$^{-1}$ and accretion to happen simultaneously \citep[see e.g.][]{gabor16}, but this mechanism might only be relevant for the brightest AGN at redshifts above $\sim$2. This is, to be sure, the period of the fastest SMBH growth, at an epoch when the major mergers rate is predicted to reach a maximum \citep[e.g.][]{hopkins10}.

Studying the incidence of very high star formation rates (SFRs) in quasar hosts, \citealt{pitchford16} (hereafter P16) recently reported on SDSS quasars with bolometric \Lagn \, in the range $10^{44.8}$ -- 10$^{47.4}$ erg s$^{-1}$ having far-infrared (FIR) luminosities between 10$^{12}$  and 10$^{14}$\Lsun. The contribution of the AGN to the FIR was estimated by means of Spectral Energy Distribution (SED) fitting, using the AGN models by \cite{feltre12}. The AGN contribution, typically reaching 20 per cent \citep{hatzimi10}, was then removed and the SFRs were computed based on the remaining FIR luminosity, attributed to star formation (for details see P16). The SFRs were found to reach up to several thousand \Msun yr$^{-1}$, thus placing the hosts of these type 1 quasars among the most luminous starburst hosts observed. Such SFRs are extraordinarily high, given that the average values for the remaining $\sim$95 per cent of the SDSS quasar population, as derived by \cite{harris16} for quasars at $2 < z < 3$ using stacking, were factors of 3-10 lower. While the majority of star-forming systems lie on a `main sequence', a tight correlation between the SFRs and stellar masses, extending from the local universe all the way to a redshift of 4.5 and possibly beyond \citep[see][and references therein]{speagle14},  
the very high SFRs alone are enough to place the quasars in P16 above the main sequence, with both accretion and star formation triggered, perhaps, by major mergers.

Such high SFRs are, however, extremely challenging to understand in terms of cosmological models for galaxy assembly. The issue for the models is that neither mergers nor cold accretion could easily produce such high SFRs; mergers because they cannot channel enough gas to the centres of haloes \citep[e.g.][]{narayanan10}, and cold accretion because massive haloes inhibit the gas flow on to central galaxies via shock heating \citep[e.g.][]{birnboim07}. Moreover, such extreme SFRs may imply {\itshape surface} SFR densities that exceed those of `maximum intensity' starbursts \citep{elmegreen99}, in which radiation pressure overcomes gravitation, thus self regulating the SFR \citep[e.g.][]{scoville01}.

At the same time, there is also some uncertainty over how real these SFRs are. The reason for this is that the spatial resolution of SPIRE, 18\pp \, FWHM at 250 \micron \, \citep{griffin10}, is rather poor, meaning that there could exist multiple FIR-bright components within the SPIRE beam. Such multiplicity in the detections of the quasars in P16 could lower the SFRs in the hosts, bringing the SFRs in line with those in \cite{harris16}, supporting at the same time the scenario that claims the brightest AGN to be triggered by major mergers, if the various counterparts are physical rather than chance associations. Interferometric observations \citep[e.g.][]{ivison12,hodge13,bussmann15,trakhtenbrot17} as well as simulated observations \citep{hayward13,cowley15} have shown that a significant fraction of single-dish sub-mm or {\it Herschel} sources are blends of multiple galaxies, but the rate and nature of
this multiplicity (physical versus chance associations) is still unclear. \cite{michalowski17}, on the other hand, using ALMA band 7 follow up observations of SCUBA sources from \cite{simpson15}  only find 15-20 per cent of bright SCUBA sources (with 850 \micron \, fluxes above 4 mJy) in the COSMOS field to be affected by multiplicity. Likewise, \cite{hill18} observed the brightest  SCUBA-2 Cosmology Legacy Survey sources with the Submillimetre Array (SMA) at 860 \micron \, and reported an upper limit of 15 per cent multiplicity in sources with two or more counterparts with flux ratios close to 1.

In the present work we discuss multiplicity rates in the sub-mm counterparts of 28 among the FIR-brightest quasars in the P16 catalogue, using recent ALMA band 7 continuum observations carried out with the Atacama Compact Array (ACA). The goals of this study are to a) determine whether one or more components lie within the {\it Herschel}/SPIRE beam and b) address the implications of the derived multiplicities in the scenario where major mergers are triggers of concomitant AGN activity and intense star formation. The quasar sample and ACA data are described in Sec. \ref{sec:sample} and \ref{sec:data}, respectively. Our findings on the ACA counterparts and their properties as well as the limitations of the data based on simulations are presented in Sec. \ref{sec:results}. Finally, in Sec. \ref{sec:discuss} we conclude this work with an in depth discussion of our results on multiplicities and the role of mergers.

\section{The sample of high-redshift infrared-bright quasars}\label{sec:sample}

Our targets are drawn from the P16 {\it Herschel}/SDSS quasar sample, that consists of all spectroscopically confirmed SDSS quasars from data releases 7 \citep[DR7;][]{schneider10} and 10 \citep[DR10;][]{paris14}, lying in some of the largest fields ever observed by {\it Herschel}. More precisely, these are the large {\it Herschel} Multi-tiered Extragalactic Survey \citep[HerMES;][]{oliver12} fields, the HerMES Large Mode Survey (HeLMS; \citealt{oliver12}, P16) and in the {\it Herschel} Stripe 82 Survey (HerS; \citealt{viero14}) fields. The 513 quasars in the P16 sample span the redshift range between 0.2 and 4.6 and they were individually detected by SPIRE at 250 \micron \, at or above 3$\sigma$. Their FIR luminosities, L$_{\textrm {FIR}}$, and SFRs were estimated by applying
a two-component (AGN+starburst) fit \citep[][P16]{hatzimi08,hatzimi09,hatzimi10} to their optical-to-FIR SEDs.

We selected a random sub-sample of 28 quasars among the brightest 500 \micron \, emitters (S$_{500} \ge 10$ mJy) at $z>2$, with individual detections in all three SPIRE bands (250, 350 and 500 \micron). For scheduling purposes, i.e. to avoid shadowing problems due to the proximity of the ACA antennas, we selected targets that lie at declinations below +3${\deg}$, in the HeLMS and HerS fields. 
The sample is presented in Tab. \ref{tab:sample} with IDs, SPIRE coordinates and fluxes, redshifts as well as the separation between the SDSS and SPIRE centroids. To the best of our knowledge, they are not lensed and they do not comply with any of the selection criteria of the SDSS Quasar Lens Search \citep[][and references therein]{inada12}. 

The $g$-band magnitudes of our targets range from 17.8 to the SDSS DR10 limit of 22.0. Their \Lagn \, range between 10$^{45.8}$ and 10$^{46.9}$ erg s$^{-1}$ with a median of 10$^{46.6}$ erg s$^{-1}$, \Mbh \, range between 10$^8$ and 7$\times 10^9$ \Msun, and their Eddington ratios, \Lagn/L$_{\textrm {Edd}}$, between 0.02 and 1 (all these quantities have been calculated for the master sample of 513 quasars as described in P16). In other words, they are representative of the SDSS quasar population at $2 \le z \le 4$. Based on the SED-fitting presented in P16, their  L$_{\textrm {FIR}}$ lie between 10$^{12.8}$ and 10$^{13.8}$ \Lsun \, and their SFRs between 1000 and 4000 \Msun \, yr$^{-1}$. Seven of them are high-ionisation broad absorption line (BAL) quasars identified in the SDSS DR12 BAL quasar catalogue, indicated in the ID column, with strong absorption blueward the C\begin{scriptsize}{IV}\end{scriptsize} 
line at 1549\textup{\AA}.

\begin{table*}
\begin{center}
\caption{FIR-bright quasar sample. From left to right, the table shows the ID with an indication whether the object is classified as a BAL quasar, optical (SDSS) and FIR ({\it Herschel}-SPIRE 250 \micron) coordinates of the source and their separation in arcsec; the SDSS redshift ($z$); and the SPIRE fluxes at 250, 350 and 500 \micron \, in mJy.}
\label{tab:sample}
\begin{tabular}{lccccccccc}
\hline
  \multicolumn{1}{c}{ID} &
  \multicolumn{2}{c}{SDSS} &
  \multicolumn{2}{c}{SPIRE 250 \micron} &
  \multicolumn{1}{c}{Separation} &
  \multicolumn{1}{c}{$z$} &
  \multicolumn{3}{c}{SPIRE fluxes [mJy]} \\
%  \multicolumn{1}{c}{Comments} \\
  & RA & Dec & RA & Dec & [$^{\prime}$$^{\prime}$]& & S$_{250}$ & S$_{350}$ & S$_{500}$ \\
\hline
J000746 & 00:07:46.93 & +00:15:43.0 & 00:07:46.93 & +00:15:42.9 & 0.13 & 2.479 & 54.47$\pm$11.8 & 43.26 $\pm$11.9 & 25.18$\pm$13.3\\
J001121 & 00:11:21.87 & --00:09:18.6 & 00:11:22.00 & -00:09:18.2 & 2.08 & 3.009 & 66.65$\pm$11.8 & 55.25$\pm$11.9 & 39.14$\pm$14.0 \\
J001401 & 00:14:01.09 & --01:06:07.2 & 00:14:01.17 & -01:06:10.9 & 3.91 & 2.103 & 52.27$\pm$11.7 & 61.91$\pm$11.9 & 45.56$\pm$13.3 \\
J003011 & 00:30:11.77 & +00:47:50.0 & 00:30:11.63 & +00:47:49.6 & 2.04 & 3.118 & 38.54$\pm$11.6 & 51.21$\pm$12.0 & 38.89$\pm$13.9  \\
J004440 & 00:44:40.50 & +01:03:06.4 & 00:44:40.56 & +01:03:07.8 & 1.64 & 3.288 & 39.51$\pm$11.7 & 27.14$\pm$12.0 & 39.80$\pm$13.8 \\
J010315 & 01:03:15.69 & +00:35:24.2 & 01:03:15.80 & +00:35:26.4 & 2.73 & 2.070 & 59.81$\pm$10.1 & 50.36$\pm$10.3 & 37.38$\pm$10.8 \\
J010524 & 01:05:24.40 & --00:25:27.1 & 01:05:24.44 & -00:25:29.3 & 2.22 & 3.529 & 40.31$\pm$10.1 & 52.31$\pm$10.4 & 32.37$\pm$11.2 \\
J010752 & 01:07:52.52 & +01:23:37.0 & 01:07:52.36 & +01:23:35.8 & 2.78 & 3.525 & 47.02$\pm$11.2 & 45.01$\pm$10.8 & 34.04$\pm$12.0 \\
J011709 & 01:17:09.57 & +00:05:23.9 & 01:17:09.36 & +00:05:24.5 & 3.28 & 2.673 & 44.78$\pm$10.5 & 48.32$\pm$10.1 & 23.60$\pm$10.7 \\
J012700$^{\rm \, (BAL)}$ & 01:27:00.69 & --00:45:59.2 & 01:27:00.78 & -00:46:00.4 & 1.73 & 4.105 & 51.98$\pm$10.0 & 63.39$\pm$ 10.1 & 35.46$\pm$ 10.8 \\
J012836 & 01:28:36.38 & +00:49:33.4 & 01:28:36.26 & +00:49:32.7 & 1.98 & 2.214 & 64.06$\pm$10.3 & 61.53$\pm$10.4 & 54.40$\pm$10.9 \\
J012845 & 01:28:45.99 & +00:38:42.9 & 01:28:46.06 & +00:38:40.4 & 2.78 & 2.758 & 32.06$\pm$10.4 & 20.28$\pm$10.1 & 25.56$\pm$10.8 \\
J013814 & 01:38:14.54 & +00:00:03.5 & 01:38:14.69 & +00:00:03.1 & 2.37& 2.153 & 65.41$\pm$11.2 & 56.95$\pm$10.9 & 30.00$\pm$11.5 \\
J014012$^{\rm \, (BAL)}$ & 01:40:12.82 & +00:28:58.0 & 01:40:12.81 & +00:28:57.7 & 0.26 & 2.728 & 37.90$\pm$10.2 & 34.13$\pm$10.2 & 21.47$\pm$10.8 \\
J014555 & 01:45:55.58 & --00:31:25.9 & 01:45:55.79 & -00:31:27.9 & 3.74 & 2.319 & 44.27$\pm$10.9 & 40.52$\pm$11.0 & 12.83$\pm$12.2 \\
J014822 & 01:48:22.70 & --00:27:12.7 & 01:48:22.67 & -00:27:13.8 & 1.16 & 2.349 & 56.67$\pm$10.1 & 53.13$\pm$10.1 & 24.67$\pm$11.0 \\
J015017$^{\rm \, (BAL)}$ & 01:50:17.71 & +00:29:02.4 & 01:50:17.85 & +00:29:04.7 & 3.05 & 3.000 & 66.21$\pm$10.1 & 83.77$\pm$10.2 & 60.77$\pm$11.0 \\
J020337$^{\rm \, (BAL)}$ & 02:03:37.23 & +00:44:46.6 & 02:03:37.26 & +00:44:46.2 & 0.59 & 2.258 & 46.08$\pm$10.3 & 46.81$\pm$ 9.97 & 46.29$\pm$11.0 \\
J020947 & 02:09:47.11 & +00:42:26.1 & 02:09:47.02 & +00:42:22.9 & 3.55 & 2.350 & 47.47$\pm$10.9 & 65.01$\pm$11.1 & 41.04$\pm$11.8 \\
J021218 & 02:12:18.50 & +00:44:55.6 & 02:12:18.62 & +00:44:56.5 & 2.08 & 2.898 & 68.14$\pm$10.6 & 77.50$\pm$10.2 & 68.74$\pm$11.0 \\
J233456 & 23:34:56.91 & --00:03:48.5 & 23:34:57.03 & -00:03:46.4 & 2.73 & 2.394 & 50.07$\pm$12.1 & 37.34$\pm$11.9 & 12.26$\pm$13.4 \\
J233600 & 23:36:00.37 & --01:50:38.6 & 23:36:00.18 & -01:50:35.2 & 4.42 & 3.174 & 52.80$\pm$12.2 & 45.93$\pm$12.0 & 34.77$\pm$13.8 \\
J233846 & 23:38:46.87 & +00:32:15.1 & 23:38:46.87 & +00:32:15.3 & 0.13 & 2.099 & 83.27$\pm$11.7 & 70.27$\pm$11.8 & 61.65$\pm$13.3 \\
J233924 & 23:39:24.66 & +00:43:56.1 & 23:39:24.68 & +00:43:55.0 & 1.13 & 2.439 & 44.48$\pm$11.8 & 32.58$\pm$12.1 & 10.00$\pm$13.8 \\
J234812$^{\rm \, (BAL)}$ & 23:48:12.99 & --03:15:01.4 & 23:48:12.79 & -03:15:04.9 & 4.61 & 2.719 & 40.35$\pm$11.7 & 50.48$\pm$11.6 & 36.75$\pm$13.5 \\
J235238$^{\rm \, (BAL)}$ & 23:52:38.09 & +01:05:52.3 & 23:52:38.07 & +01:05:54.1 & 1.81 & 2.153 & 65.75$\pm$12.2 & 65.35$\pm$11.8 & 45.44$\pm$13.4 \\
J235859$^{\rm \, (BAL)}$ & 23:58:59.51 & +02:08:47.5 & 23:58:59.27 & +02:08:45.6 & 4.03 & 2.910 & 57.88$\pm$11.8 & 44.99$\pm$12.0 & 42.39$\pm$13.3 \\
J235944 & 23:59:44.94 & +02:29:06.9 & 23:59:44.77 & +02:29:03.8 & 4.12 & 2.258 & 51.00$\pm$12.3 & 40.51$\pm$12.0 & 30.74$\pm$13.7 \\
\hline\end{tabular}
\end{center}
\end{table*}

\section{The ACA data}\label{sec:data}

The sample of 28 quasars was divided into four groups by the clustering algorithm of the ALMA Observing Tool, that effectively translate into four individual Scheduling Blocks (SBs) at the time of observations. The observations were carried out as part of project 2016.2.00060.S (PI: Hatziminaoglou) between July 4 and 29, 2017 in single continuum spectral setup at a representative frequency of 350 GHz, with on time on source ranging from 36 to 43 minutes, with synthesised beams $\sim$4\ppdot3 $\times$ 3\ppdot0 (varying slightly among the various SBs, see right-most column in Tab. \ref{tab:separations}). The beam was centred on the SDSS positions. 

\begin{figure*}
%\captionsetup[subfigure]{labelformat=empty}
\includegraphics[width = 5cm]{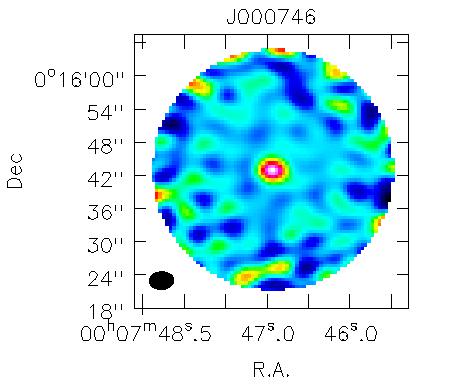}
\includegraphics[width = 5cm]{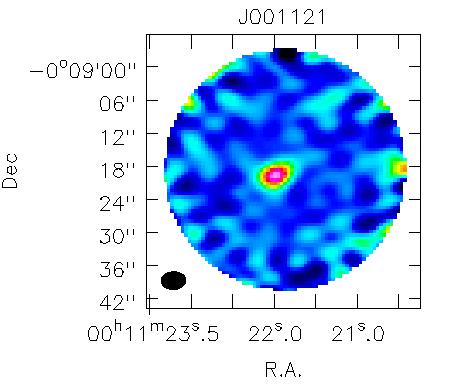}
\includegraphics[width = 5cm]{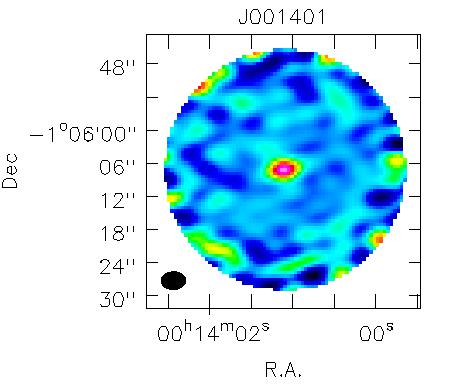} \\
\includegraphics[width = 5cm]{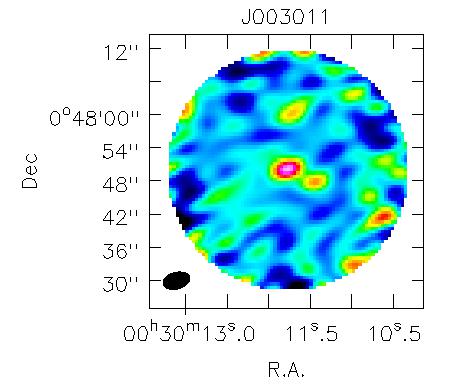}
\includegraphics[width = 5cm]{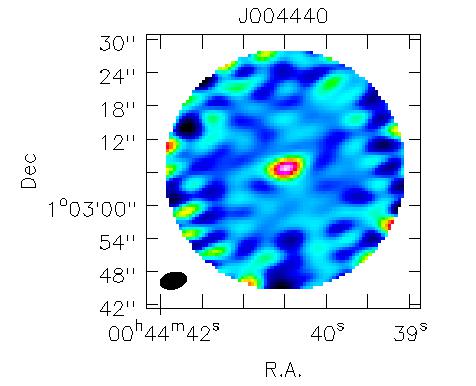}
\includegraphics[width = 5cm]{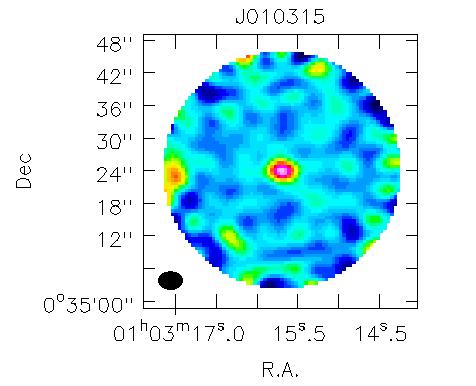} \\ 
\includegraphics[width = 5cm]{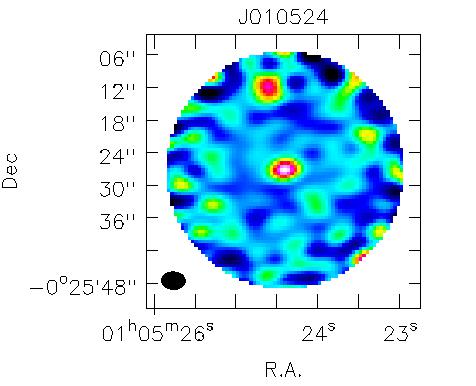}
\includegraphics[width = 5cm]{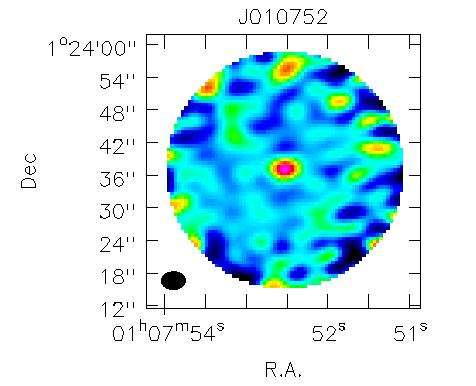} 
\includegraphics[width = 5cm]{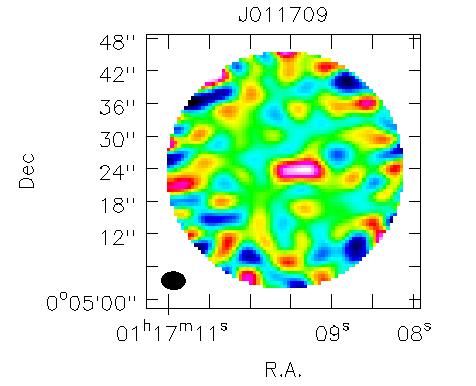} \\
\includegraphics[width = 5cm]{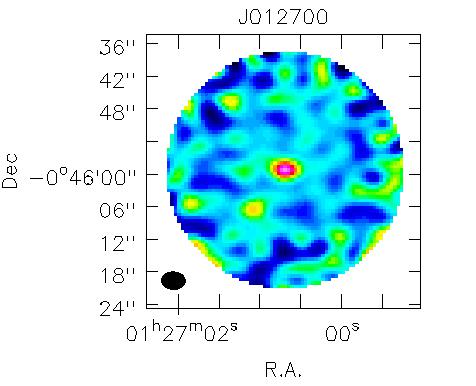}  
\includegraphics[width = 5cm]{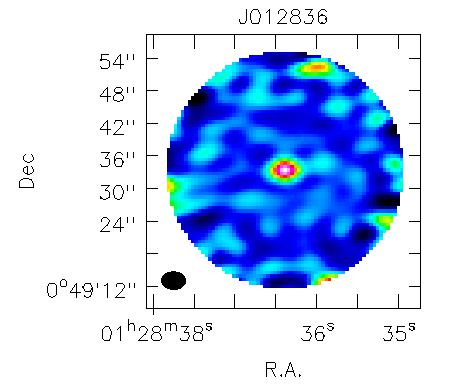} 
\includegraphics[width = 5cm]{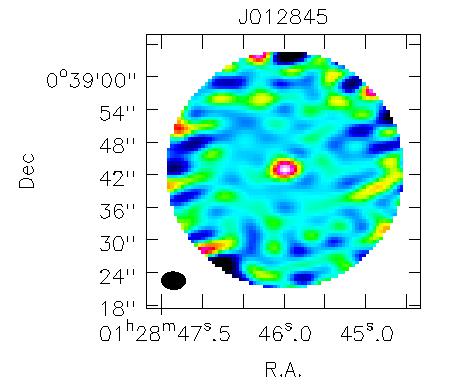} \\
\includegraphics[width = 5cm]{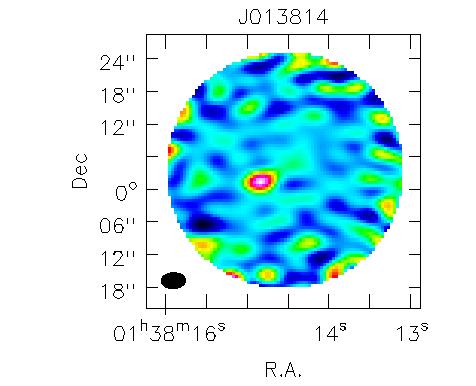} 
\includegraphics[width = 5cm]{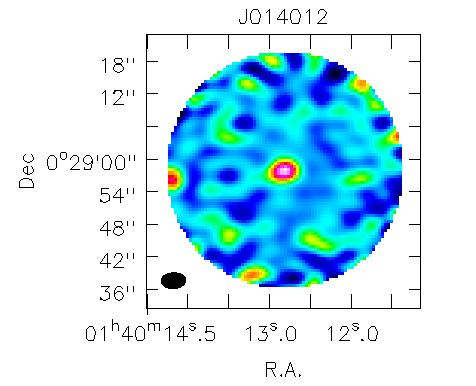} 
\includegraphics[width = 5cm]{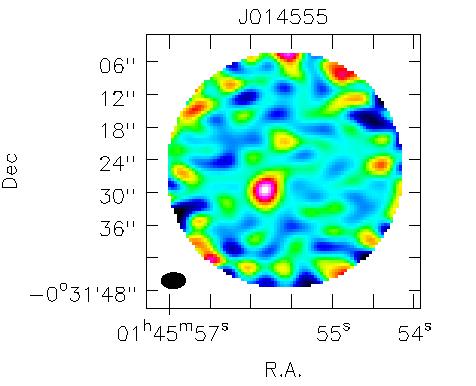} \\
\caption{870 \micron \, maps of the 28 {\it Herschel}/SDSS quasars of our sample (North is up and East is left). The beam is shown at the bottom left of each image. The diameter of the cutouts is 43$^{\prime\prime}$.}
\label{fig:cutouts}
\end{figure*}

The delivered data were pipeline-reduced using CASA version 4.7.2. We re-calibrated and re-imaged the data using the pipeline in CASA version 5.1, using the flagging files, antenna position corrections and flux files (flux.csv) provided by the ALMA observatory quality assurance process. The final images were corrected for the primary beam, though all targets are located within 6\pp \, from the centre of the ALMA pointings. Fig.  \ref{fig:cutouts} shows the ACA 870 \micron \,  continuum maps for the full sample.

\setcounter{figure}{0}
\renewcommand{\thefigure}{\arabic{figure} (continued)}

\begin{figure*}
%\captionsetup[subfigure]{labelformat=empty}
\includegraphics[width = 5cm]{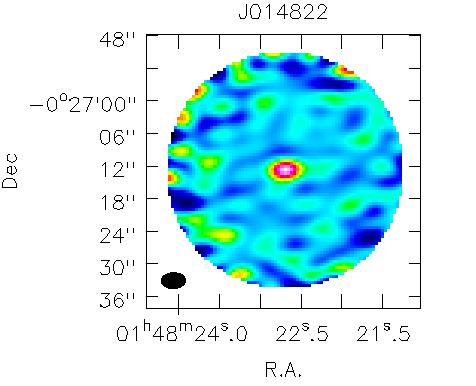} 
\includegraphics[width = 5cm]{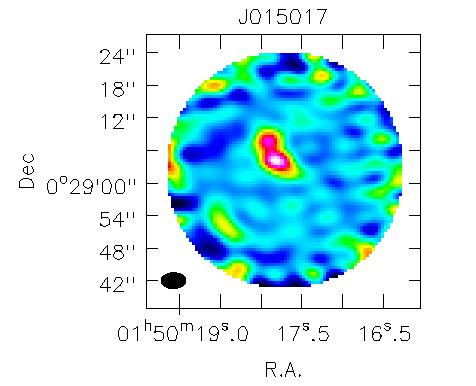} 
\includegraphics[width = 5cm]{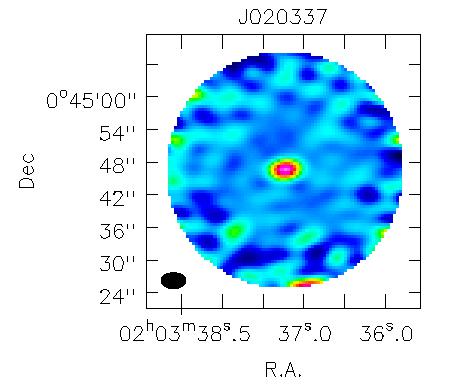} \\
\includegraphics[width = 5cm]{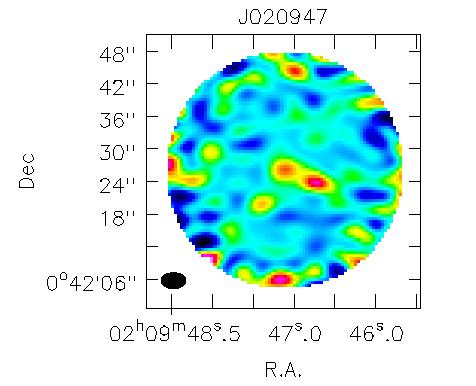} 
\includegraphics[width = 5cm]{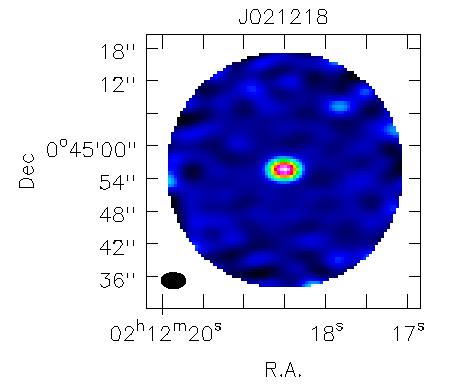} 
\includegraphics[width = 5cm]{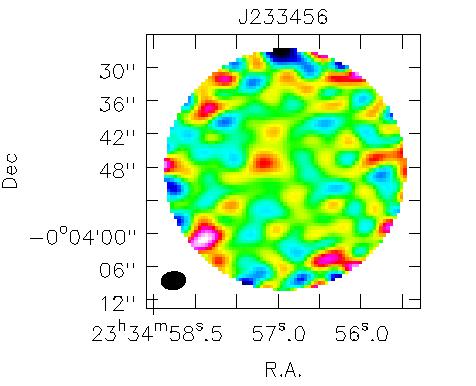} \\
\includegraphics[width = 5cm]{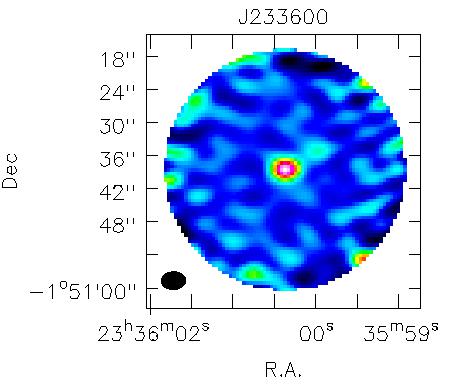} 
\includegraphics[width = 5cm]{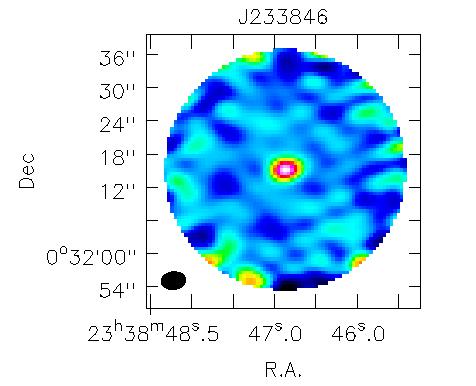}
\includegraphics[width = 5cm]{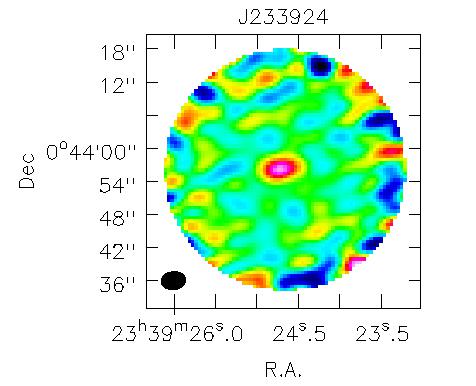} \\
\includegraphics[width = 5cm]{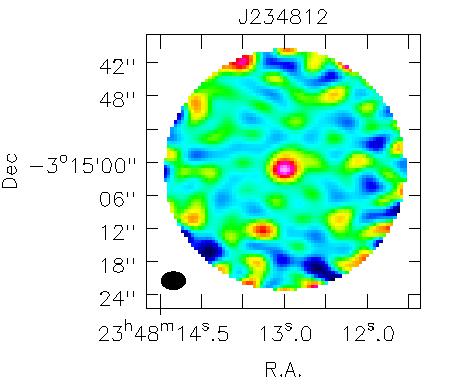} 
\includegraphics[width = 5cm]{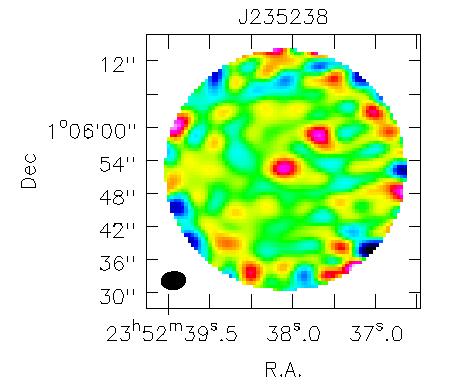} 
\includegraphics[width = 5cm]{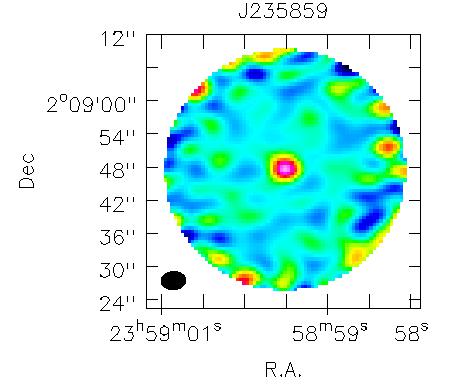} \\
\includegraphics[width = 5cm]{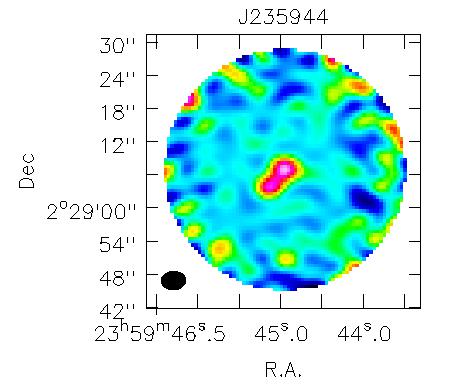} 
\caption{}
\end{figure*}

\section{Results}\label{sec:results}

Taking into account the SPIRE 250 \micron \, beam of 18\pp \, FWHM and the {\it Herschel} pointing accuracy of $<$ 2\pp, we look for counterparts within (18\pp/2)+2\pp = 11\pp \, from the SPIRE coordinates.

All of the ACA maps show at least one detection at 870$\,\mu$m within 11\pp \, of the SPIRE coordinates. Source J233456, the source with the lowest ACA detection level, is detected at 3.5$\sigma$. All other sources are detected at or above 5$\sigma$, with 18 detections at or above 9$\sigma$. The counterparts, their coordinates, fluxes, distances from the SDSS and SPIRE 250 \micron \, centroids and the restored beam for each pointing are listed in Tab. \ref{tab:separations}. Of the 28 SDSS quasars, 19 have unique ALMA counterparts, while six and three sources have a second and third counterparts within 11\pp, respectively. The few sources lying beyond $\sim$ 11\pp \, from the SPIRE coordinates are not included here, as they are less likely to be associated with the SPIRE detections \citep[e.g.][]{hayward18}, but they are discussed in Appendix \ref{sec:other} and are shown in Tab. \ref{tab:other}.

We define the primary ACA counterpart to each SDSS source as either the unique ACA counterpart, or the closest ACA counterpart in case of multiplicity. Fig. \ref{fig:s870histo} shows the distribution of S$_{870}$, for all sources within 11\pp \, from the SPIRE coordinates listed in Tab. \ref{tab:separations} (grey area), for the primary counterparts (black solid histogram) the secondary counterparts (long red dashed spikes) and the sum per pointing (short green dashed spikes). The full S$_{870}$ distribution is bimodal, with a peak at $\sim$4 mJy and another one at $\sim$8 mJy and a tail extending towards brighter fluxes. The brighter peak is almost entirely due to the primary components while the fainter peak is composed mainly of secondary counterparts.

\setcounter{figure}{1}
\renewcommand{\thefigure}{\arabic{figure}}

\begin{figure}
\centering\includegraphics[width=0.45\textwidth]{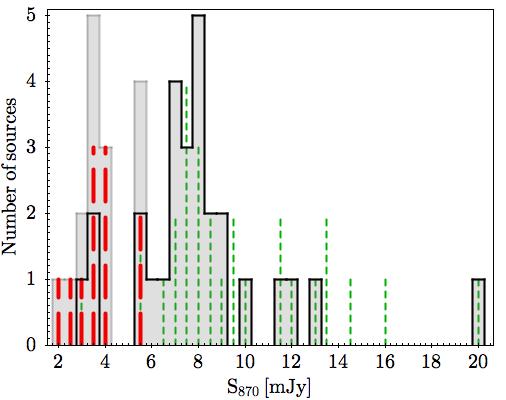}
\caption{The S$_{870}$ flux density distribution of the sample: the grey area shows all sources within 11$^{\prime\prime}$ from the SPIRE coordinates, the black solid histogram indicates the primary counterparts, the long red dashed spikes the secondary components within 11$^{\prime\prime}$ and the short green dashed spikes the sum per pointing.}
\label{fig:s870histo}
\end{figure}

\begin{table*}
\begin{center}
\caption{ACA counterparts within 11$^{\prime}$$^{\prime}$ from the SPIRE 250 \micron \, positions, 870 \micron \, fluxes, and synthesised beam as measured on the ALMA images.}
\label{tab:separations}
\begin{tabular}{lccccrc}
\hline
  \multicolumn{1}{c}{ID} &
  \multicolumn{2}{c}{ACA} &
  \multicolumn{2}{c}{Separation from} & 
  \multicolumn{1}{c}{870 \micron \, flux} &
  \multicolumn{1}{c}{Synthesised beam} \\
  & RA & Dec & SDSS [$^{\prime}$$^{\prime}$] & SPIRE [$^{\prime}$$^{\prime}$] & [mJy] & \\
\hline
J000746 & 00:07:46.93 & +00:15:42.9 & 0.14 & 0.1 & 6.69$\pm$0.30 & 4\ppdot40 $\times$ 3\ppdot17\\
J001121 & 00:11:21.98 & --00:09:19.6 & 1.94  &1.43 & 11.44$\pm$0.67 & 4\ppdot40 $\times$ 3\ppdot16 \\
J001401 & 00:14:01.11 & --01:06:07.0 & 0.33 & 4.00 & 7.42$\pm$0.42 & 4\ppdot39 $\times$ 3\ppdot17\\
J003011 & 00:30:11.77 & +00:47:50.1 & 0.10 & 2.16 & 7.22$\pm$0.50 & 4\ppdot89 $\times$ 2\ppdot96  \\ 
& 00:30:11.45 & +00:47:47.9 & 5.37 & 3.14 & 3.31$\pm$0.30 \\
& 00:30:11.70 & +00:48:00.6 & 10.35 & 10.75 & 4.09$\pm$0.51 & \\ 
J004440 & 00:44:40.48 & +01:03:06.7 & 0.37 & 1.63 & 8.99$\pm$0.48 & 4\ppdot82 $\times$ 2\ppdot95\\
J010315 & 01:03:15.69 & +00:35:24.0 & 0.07 & 2.83 & 6.85$\pm$0.21 & 4\ppdot35 $\times$ 3\ppdot20\\
J010524 & 01:05:24.40 & --00:25:27.0 & 0.06 & 2.28 & 8.00$\pm$0.21 & 4\ppdot29 $\times$ 3\ppdot20\\
J010752 & 01:07:52.52 & +01:23:37.1 & 0.13 & 2.73 & 5.54$\pm$0.24 & 4\ppdot35 $\times$ 3\ppdot23\\
J011709 & 01:17:09.38 & +00:05:23.7 & 2.90 & 0.85 & 8.23$\pm$0.81 & 4\ppdot32 $\times$ 3\ppdot21\\
J012700 & 01:27:00.69 & --00:45:59.1 & 0.12 & 1.87 & 5.97$\pm$0.34 & 4\ppdot28$\times$ 3\ppdot19\\
& 01:27:01.06 & --00:46:06.3 & 8.98 & 7.24 & 3.43$\pm$0.14 \\
J012836 & 01:28:36.38 & +00:49:33.4 & 0.07 & 1.93 & 9.85$\pm$0.29 & 4\ppdot36 $\times$ 3\ppdot20\\
J012845 & 01:28:45.99 & +00:38:43.0 & 0.05 & 2.71 & 5.35$\pm$0.40 & 4\ppdot48 $\times$ 3\ppdot24\\
J013814 & 01:38:14.83 & +00:00:01.5 & 4.82 & 2.64 & 7.30$\pm$0.24 & 4\ppdot45 $\times$ 2\ppdot89\\
J014012 & 01:40:12.82 & +00:28:57.7 & 0.27 & 0.15 & 7.83$\pm$0.29 & 4\ppdot46 $\times$ 2\ppdot88\\
J014555 & 01:45:55.81 & --00:31:29.5 & 5.99 & 1.63 & 9.13$\pm$0.46 & 4\ppdot43 $\times$ 2\ppdot88\\ 
& 01:45:55.60 & --00:31:20.6 & 5.18 & 7.74 & 2.37$\pm$0.26\\
J014822 & 01:48:22.70 & --00:27:12.7 & 0.03 & 1.19 & 8.38$\pm$0.31 & 4\ppdot43 $\times$ 2\ppdot88\\
J015017 & 01:50:17.80 & +00:29:04.0 & 2.02 & 1.03 & 8.16$\pm$0.72 & 4\ppdot50 $\times$ 2\ppdot88\\ 
& 01:50:17.91 & +00:29:07.6 & 5.94 & 3.04 & 5.46$\pm$0.82 \\
J020337 & 02:03:37.23 & +00:44:46.6 & 0.04 & 0.60 & 8.39$\pm$0.24 & 4\ppdot52 $\times$ 2\ppdot89\\
J020947 & 02:09:47.10 & +00:42:26.0 & 0.22 & 3.32 & 3.53$\pm$0.12 & 4\ppdot53 $\times$ 2\ppdot90\\ 
& 02:09:46.73 & +00:42:23.8 & 6.18 & 4.44 & 3.96$\pm$0.10 \\ 
& 02:09:47.38 & +00:42:19.8 & 7.51 & 6.23 & 2.03$\pm$0.10 \\
J021218 & 02:12:18.50 & +00:44:55.6 & 0.08 & 2.01 & 20.15$\pm$0.20 & 4\ppdot52 $\times$ 2\ppdot92\\
J233456 & 23:34:57.16 & --00:03:47.3 & 3.92 & 2.11 & 2.90$\pm$0.11 & 4\ppdot39 $\times$ 3\ppdot20\\
J233600& 23:36:00.36 & --01:50:38.4 & 0.23 & 4.19 & 12.78$\pm$0.22 & 4\ppdot37$\times$ 3\ppdot18\\
 & 23:35:59.97 & --01:50:35.1 & 6.66 & 2.85 & 3.35$\pm$0.10 \\ 
J233846 & 23:38:46.87 & +00:32:15.2 & 0.06 & 0.1 & 12.13$\pm$0.42 & 4\ppdot41 $\times$ 3\ppdot18 \\
J233924 & 23:39:24.72 & +00:43:56.3 & 0.92 & 1.43 & 7.14$\pm$0.19 & 4\ppdot39 $\times$ 3\ppdot21\\
J234812 & 23:48:12.99 & --03:15:01.1 & 0.38 & 4.92 & 6.97$\pm$0.33 & 4\ppdot36 $\times$ 3\ppdot16\\
 & 23:48:12.32 & --03:15:10.5 & 9.12 & 5.62 & 2.81$\pm$0.15 \\ 
 & 23:48:13.24 & --03:15:12.3 & 11.54 & 10.01 & 3.05$\pm$0.39 \\
J235238 & 23:52:38.10 & +01:05:52.6 & 0.34 & 1.57 & 3.75$\pm$0.13 & 4\ppdot40 $\times$ 3\ppdot21 \\
 & 23:52:37.67 & +01:05:58.5 & 8.77 & 7.44 & 3.92$\pm$0.30 \\ 
J235859 & 23:58:59.51 & +02:08:47.8 & 0.29 & 4.22 & 7.92$\pm$0.24 & 4\ppdot41 $\times$ 3\ppdot25\\
J235944 & 23:59:44.93 & +02:29:07.0 & 0.22 & 4.00 & 7.93$\pm$0.92 &4\ppdot39 $\times$ 3\ppdot29 \\
& 23:59:45.10 & +02:29:04.4 & 3.42 & 4.98 & 5.40$\pm$0.83& \\
\hline\end{tabular}
\end{center}
\end{table*}

\subsection{Astrometry}\label{sec:astro}
\begin{figure}
\centering\includegraphics[width=0.45\textwidth]{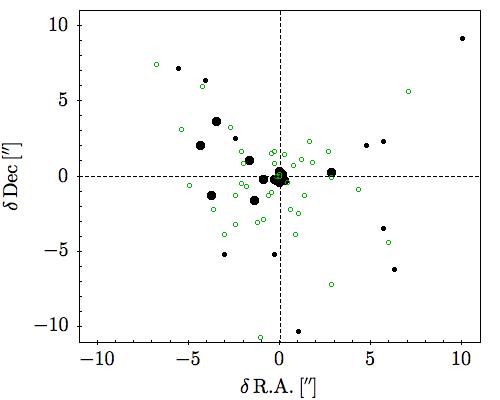}
\caption{Astrometric offset between the ACA and SDSS (black filled symbols), and between ACA and SPIRE (green open symbols) for all counterparts. Primary counterparts (i.e. unique counterparts or those lying closer to the SDSS position in case of multiple counterparts) are shown with larger symbols.}
\label{fig:radec}
\end{figure}

We present the astrometric offsets between the ACA, SDSS and SPIRE sources in Fig. \ref{fig:radec}. Of the 19 objects with unique ACA counterparts, the ACA coordinates of 15 of them are within 1\pp \, of the SDSS coordinates. For objects J001121, J011709, J013814 and J233456 the coordinates of the ACA counterpart are closer to those of the SPIRE centroid than the SDSS quasar by 0\ppdot5, 2\ppdot1, 2\ppdot2 and 1\ppdot8, respectively. Moreover, J011709 is a blended source in the ACA map but the resolution does not allow us to separate the components. 

Of the nine sources with more than one ACA counterparts, seven have their brightest ACA counterpart within 1\pp \, of the SDSS coordinates.  
For the objects with offsets greater than 1\pp \, between the ACA and SDSS positions, we checked the SDSS coordinates against GAIA and UKIDSS. Indeed, the GAIA coordinates of J013814 are identical to the SDSS ones. For the remaining objects, which do not fall into the footprint of the GAIA DR1, the UKIDSS DR10+ coordinates coincide to within 0\ppdot5 from the SDSS coordinates. Given the astrometric accuracy of SDSS \citep{pier03} as well as that of ACA ($<$0\ppdot6; \citealt{saito12}), we deduce that the offsets are real.

\subsection{The multiplicity of the ACA sources}\label{sec:multi}

Before discussing the multiplicities found in this work, the word has to be defined in this particular framework. From the distribution of 870 \micron \, fluxes we find that for primary sources with fluxes around $\sim$ 8 mJy, corresponding to the median of the sample, the secondary counterparts have fluxes between 30 and 70 per cent that of the primary. The secondary counterpart of the brightest multiple source (J233600) has about a 25 per cent of the flux of the primary. On the other hand, the faintest source with close companions (J020947), that is also the second faintest object at 870 \micron \, in the sample, has one counterpart with half the flux and one with comparable flux. All but one single sources have fluxes very near the median of the sample and, therefore, companions would be detectable if present. In other words, at fluxes around or above the median of the sample, the ACA data have enough sensitivity to detect multiple systems with counterparts each of which contributes to at least 25 per cent of the total 870 \micron \, flux per pointing.

\begin{figure}
\centering\includegraphics[width=0.45\textwidth]{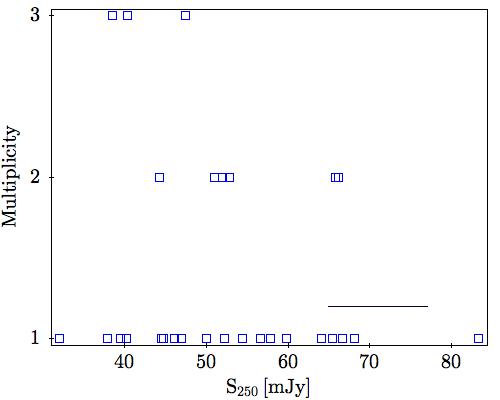}
\caption{Number of counterparts within 11$^{\prime\prime}$ from the SPIRE 250 \micron \, coordinates as a function of S$_{250}$. A typical SPIRE 250 \micron \, photometric error is indicated by the black line segment.}
\label{fig:multi500}
\end{figure}

Fig. \ref{fig:multi500} shows the multiplicity as a function of the 250 \micron \, flux, S$_{250}$, where no clear dependence is evident. We tentatively highlight that all three sources with three counterparts have S$_{250}$ values below the median S$_{250}$ of the sample, but the number of sources is too low to draw any reliable conclusions. 

\begin{figure}
\centering\includegraphics[width=0.45\textwidth]{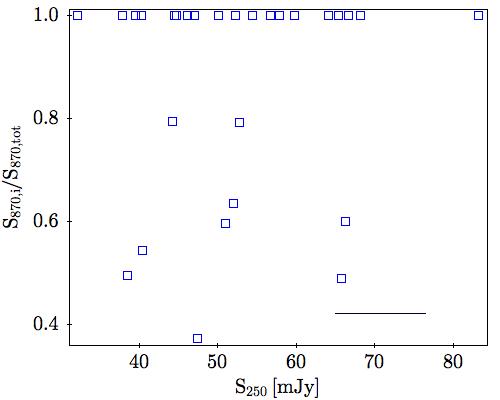}
\caption{Fraction of the 870 \micron \, flux of the brightest ALMA component, as a function of the SPIRE 250 \micron \, flux. A typical SPIRE 250 \micron \, photometric error is indicated by the black line segment.}
\label{fig:sistot}
\end{figure}

In Fig. \ref{fig:sistot}, the fraction of the total 870 \micron \, flux arising from the brightest ACA counterpart, S$_{870,{\textrm i}}$/S$_{870, {\textrm{tot}}}$, is shown as a function of the SPIRE 250 \micron \, flux, S$_{250}$. For almost all of the sources with more than one ACA counterpart, the brightest one accounts for at least half of the total S$_{870}$. This figure is analogous to Fig. 4 in \cite{scudder16}, that shows the contribution of the brightest component to the SPIRE 250 \micron \, flux, instead, again as a function of S$_{250}$. The striking difference between the two results is the large fraction of sources in our sample that have a unique ACA counterpart, even though our data would allow the detection of secondary sources contributing at or above a 25 per cent level around primary sources with the median flux of the sample.

Note that none of the secondary counterparts are visible in the SDSS or UKIDSS images (see Fig. \ref{fig:allcutouts}). The resolution of 6\ppdot1 and 6\ppdot4 in the 3.4 $\mu$m and 4.6 $\mu$m WISE bands, respectively, is larger than the separation between most of the ACA counterparts. Due to all of the above, nothing can be said about the nature of these sources.

\subsection{Limits on the multiplicity imposed by the ACA resolution}\label{sec:sim}

\begin{figure}
\includegraphics[width=0.47\textwidth]{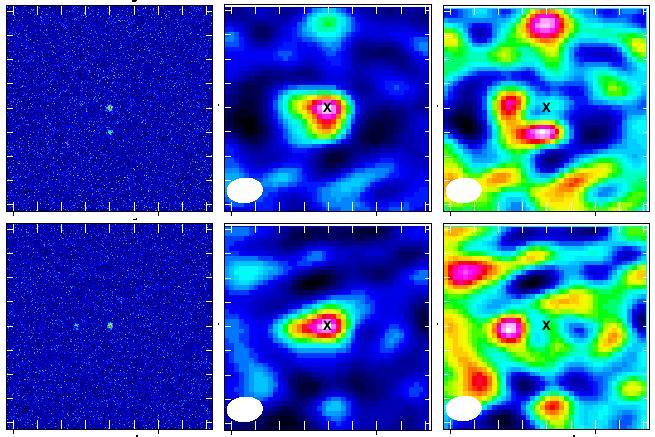}
\caption{Examples of simulated pairs of point-like sources at a distance of 3$^{\prime\prime}$ from each other (left) , smoothed to the ACA resolution (middle) and the residuals after subtracting the ACA beam (right). The black crosses mark the position of the primary source. For details, see the text. The cutouts have sizes of 24$^{\prime\prime} \times 24^{\prime\prime}$.}
\label{fig:examplesims}
\end{figure}

To investigate the quantitative limits that can be put on the multiplicity fraction we find here given the resolution of the ACA data, we produced a series of simulated point-like (Gaussian) pairs, with relative distances between 1\pp \, and 5\pp \, flux ratios between 1:1 and 3:1, lying almost along a) the major axis of the ACA beam (representing the worse case scenario) and b) the minor axis of the ACA beam (representing the best case scenario), with an original size of 0\ppdot5 (FWHM) for each of the sources. The flux of the primary source was fixed to $\sim$8 mJy (the median value of our sample) in all cases. We rebinned and smoothed the simulated images to the resolution of the ACA data. We then fit with the ACA beam to the primary (brightest and centred) component on each image and look for residuals at the position of the secondary (fainter) component. Fig. \ref{fig:examplesims} shows examples of simulated pairs (left) lying along the minor axis of the ACA beam (top) at a distance of 3\pp \, and along the major axis (bottom) at a distance of 4\pp \, from one another, with a flux ratio of 2:1, smoothed to the ACA resolution (middle) and their residual images (right). 

\begin{figure}
\includegraphics[width=0.45\textwidth]{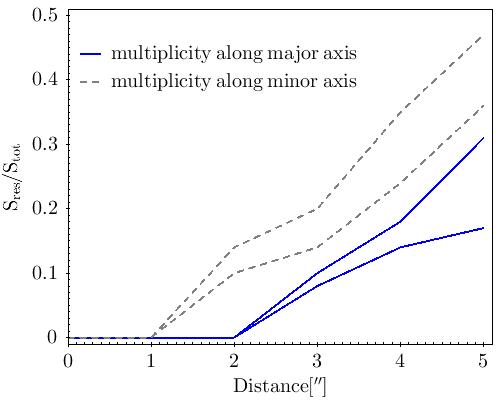}
\caption{Fraction of the residual flux at the position of the secondary source over the total flux of the system, S$_{\rm res}$/S$_{\rm tot}$, as a function of the distance between the two sources of the pair. The lower and upper curves of each set correspond to flux ratios of the primary over the secondary sources of 3:1 and 1:1, respectively.}
\label{fig:residuals}
\end{figure}

Regardless of the relative position between the two sources, the residuals at the position of the secondary source, after subtracting the ACA beam, increase with the increasing distance and with decreasing flux ratio between the two sources forming each pair. This is shown in Fig. \ref{fig:residuals}, that illustrates the fraction of the residual flux at the position of the secondary source over the total flux of the pair, S$_{\rm res}$/S$_{\rm tot}$, as a function of the distance between the two sources of the pair. The dashed grey and full blue lines correspond to sources lying along the major and minor axis of the ACA beam, respectively. In both cases, the lower and upper curves correspond to flux ratios of 3:1 and 1:1, respectively.

\begin{figure}
\centerline{
\includegraphics[width=0.265\textwidth]{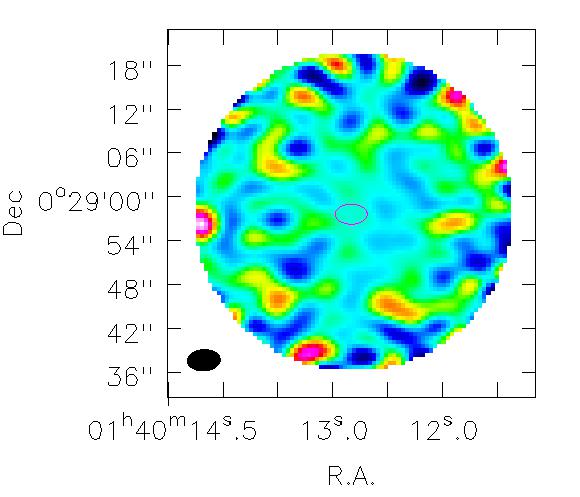}
\hskip -0.3cm
\includegraphics[width=0.255\textwidth]{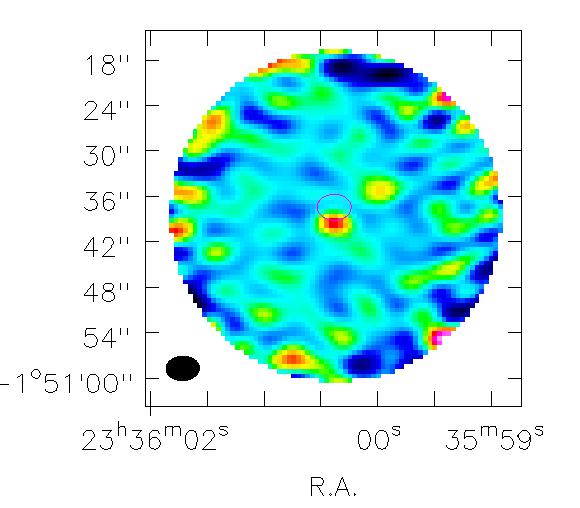}
}
\caption{Left: Map around J014012 after subtracting the ACA beam at the position of the primary source, marked by the ellipse in the centre of the image. No residual flux is detectable in the vicinity of the primary source. Right: same for J233600, with a clear residual visible at a distance of 2\ppdot5 south from the position of the primary source.}
\label{fig:j233600}
\end{figure}

As shown in these two figures, sources lying further apart than 2\pp \, along the minor axis of the ACA beam or 3\pp \, along the major axis of the beam will cause the detection on the image to look extended and no longer beam-like and will leave clear residuals at the position of the secondary source. We therefore applied the same procedure to the 28 ACA images looking for hidden counterparts inside what look like single, point-like (beam-shaped) sources. All but one sources left no residuals after subtracting the beam (for an example, see the left panel of Fig. \ref{fig:j233600}). The only exception is J233600, with a S$_{\rm res}$/S$_{\rm tot}$=0.27 at a distance of 2\ppdot5 \, to the south of the position of the primary source (i.e. lying along the minor axis of the ACA beam), shown in the right panel of Fig. \ref{fig:j233600}. This system was already among those with two counterparts and therefore this finding does not change the results on multiplicities, it does confirm, however, that this type of analysis is capable of revealing counterparts lying inside the beam, as suggested by the simulations.

\begin{figure*}
\centerline{
\includegraphics[width=0.33\textwidth]{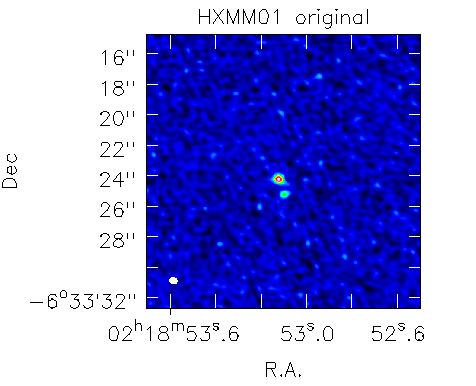}
\includegraphics[width=0.33\textwidth]{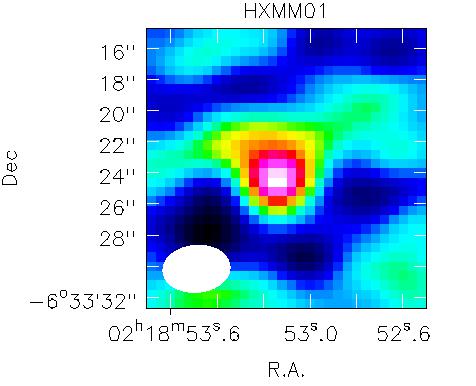}
\includegraphics[width=0.33\textwidth]{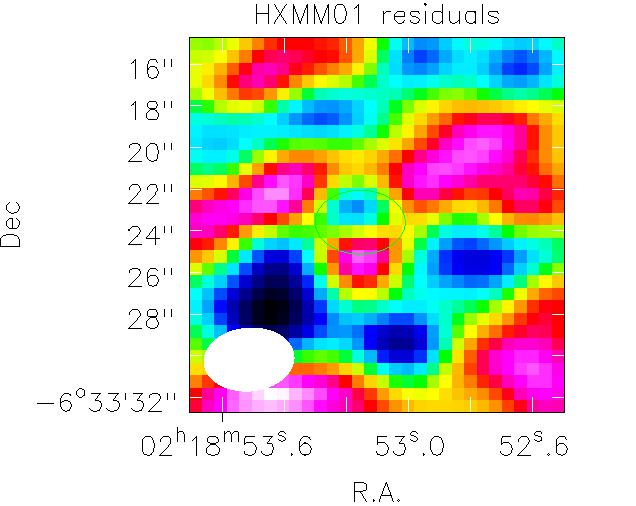}
}
\caption{Example of a double source (HXMM01) from the Bussmann et al. 2015 sample in its original form (left), smoothed with our ACA beam (middle) and the residuals after subtracting the beam at the position of the primary source (right).}
\label{fig:exampleBussmann}
\end{figure*}

Finally, we degraded ALMA band 7 continuum maps of sub-millimetre galaxies (SMGs) at 0\ppdot45 from \cite{bussmann15} to the resolution of our ACA data and compared the output images with our own data. We obtained the ALMA maps from the ALMA archive (Project 2011.1.00539.S; PI Riechers) and smoothed them to the ACA resolution, assuming a beam of 4\ppdot3 $\times$ 3\ppdot0 with a position angle (PA) of 85 degrees, representative of our ACA observations, using the CASA task {\tt imsmooth}. In all cases of multiplicity, we detected residual flux at the position of the secondary source, even for their field5 (HXMM01), composed by two source separated by 1\pp \,\, and with a flux ratio between the primary and secondary components of 3:1, albeit with S$_{\rm res}$/S$_{\rm tot}$=0.05. This particular case is shown in Fig. \ref{fig:exampleBussmann}. 

These tests indicate that the ACA sources are consistent with being point-like and that the ACA observations are capable of detecting counterparts lying within the ACA beam, separated by 2\pp \,\, to 3\pp, depending on their relative position. We are therefore quite confident that the 19 sources that look isolated and point-like on our images are indeed single sources or pairs separated by less than 2\pp -- 3\pp. This, in turn, means that the multiplicity rate reported here is a factor of two lower than \cite{bussmann15}. We will go back to the implications of this point in Sec. \ref{sec:discuss}.

As a last remark we would like to point out that four out of the seven BAL quasars in the sample are among the nine sources with more than one counterparts on the ACA images (J012700, J015017, J234812 and J235238), with the second counterpart lying, for all but J015017, further away than 7\pp \, from the SDSS quasar. For J015017 and J234812, the second counterpart lies 3\pp \, and 5\ppdot6, respectively, from the SPIRE coordinates. This suggests that BAL quasars may show a higher multiplicity rate than non-BAL quasars. However, due to the small number of sources, no firm conclusions can be drawn.

\section{Discussion}\label{sec:discuss}

Our targets were selected to have some of the most extreme SFR occurrences of any quasar, 
estimated based on their {\it Herschel}/SPIRE luminosities, after subtracting the contribution of the AGN, to reach beyond 1000 \Msun yr$^{-1}$, as discussed in P16. However, the resolution of SPIRE is coarse at 250 \micron \, (18\pp). Therefore, the questions that arise are threefold - are the FIR fluxes emerging from single sources or can they be broken down into multiple counterparts, are they, along with the associated SFRs, coming from the quasar host and what is the triggering mechanism? Using the ACA band 7 continuum observations  at 870 \micron \, of 28 FIR-bright SDSS quasars selected from P16 with redshifts between 2 and 4 , we take a step towards answering these questions.

\subsection{Multiplicity rates of sub-mm sources}

In 19/28 ($\sim70$ per cent) FIR-bright quasars the ACA observations detect a single, unresolved sub-mm counterpart (see Fig. \ref{fig:cutouts} as well as the beams in Tab. \ref{tab:separations}). From the remaining nine quasars, six (three) break into two (three) counterparts each, but, as described in Sec. \ref{sec:sim}, inside the beam of the main source of one of the double systems may be hiding another counterpart. For 15 of the single and seven of the nine multiple sources (i.e. 80 per cent in either case), the primary ACA counterpart is centred on the SDSS coordinates to within 1\pp. Let us now discuss how this multiplicity rate compares to other results on multiplicities in sub-mm sources.

Using Bayesian inference methods to produce probability distributions of the possible contributions to the observed 250 \micron \, flux for each potential component, \cite{scudder16} found that the brightest {\it Herschel} sources (S$_{250} > 45$ mJy) are typically composed of at least two and in some cases more counterparts with the brightest contributing to about 40 per cent of the total 250 \micron \, flux, while the faintest 250 \micron \, sources (30 - 45 mJy) have the majority of their flux assigned to a single bright component. Simulations by \cite{bethermin17} find a multiplicity rate comparable to \cite{scudder16}, but with an increasing fractional contribution of the brightest component to the total 250 \micron \, flux.

Out of the 28 quasars studied here, 18 have 250 \micron \, flux higher than 45 mJy. However, only six of those (or 33 per cent) have a second counterpart, with the primary source contributing from 50 to 80 per cent of the total 870 \micron \, flux. At the same time, three of the nine sources with multiple ACA counterparts have 250 \micron \, fluxes below 45 mJy, a region where according to \cite{scudder16} the SPIRE sources are not expected to break into counterparts. The present study does not, therefore, confirm the conclusions drawn by \cite{scudder16}. We point out, however, that the sample discussed in their study comprises different populations (some of which may be more prone to multiplicity effects than others), with high-redshift optically- and FIR-bright quasars most likely being a small minority. 

The single counterpart fraction reported here is also much higher than what has been reported in certain studies on SMGs. ALMA 870 \micron \, (band 7) continuum observations of 29 dusty star-forming galaxies in the HerMES fields at a 0\ppdot45 resolution \citep{bussmann15} revealed that a large fraction (80 per cent excluding lensed objects) break down into multiple counterparts. These counterparts are typically located within 2\pp of each other, with the brightest contributing to at least 35 per cent of the total 870 flux. \cite{michalowski17}, on the other hand, only report a 15-20 per cent multiplicity among bright SCUBA sources in the COSMOS field observed with ALMA at 0\ppdot3 resolution, again considering as multiple counterparts sources that contribute a minimum of 30 per cent to the total flux at 850 \micron. With our definition of multiplicity directly comparable to that in both works (i.e. counterparts each accounting for at least 25 per cent of the total 870 \micron \, flux), our results are more in line with \cite{michalowski17}.

\cite{hodge13} discuss ALMA band 7 continuum observations of sub-mm \citep[ALESS;][]{swinbank14} sources at 1\ppdot6 resolution, based on which they report a  35 -- 50 per cent multiplicity rate. This value is up to a factor two lower than that reported by \cite{bussmann15} but with a resolution of 0\ppdot45, and up to a factor of two higher than our own findings. Smoothing their images to match the observations conducted by \cite{hodge13}, \cite{bussmann15} recover a multiplicity rate of about 55 per cent. The two samples, however, have very different 870 \micron \, flux distributions, with 6 mJy median for the former and 14.9 mJy for the latter, perhaps hinting towards flux playing a role in multiplicity. For comparison, the flux distribution of our quasar sample is comparable to the ALESS sample (see 870 \micron \, flux distribution in Fig. \ref{fig:s870histo}), and so does the multiplicity rate we recover.

Finally, \cite{trakhtenbrot17} recently reported on ALMA band 7 continuum observations at 0\ppdot3 resolution of six luminous quasars at $z\sim4.8$, three FIR-bright and three FIR-faint (based on {\it Herschel}/SPIRE observations). Their study revealed companion SMGs at distances between 14 and 45 kpc for thee out of the six quasars, one FIR-bright and two FIR-faint. Though at higher redshift than our most distant sources, our sample compares to the FIR-bright objects of this study in terms of \Lagn, \Mbh \, Eddington ratios, and S$_{870}$, as well as in multiplicity rate, though the Trakhtenbrot FIR-bright sub-sample is admittedly very small to derive any statistical conclusions. Very recently, additional ALMA observations a larger sample of $z\sim4.8$ quasars carried out by the same group brought the multiplicity fraction of the total sample to 30 per cent, but the fraction changes strongly between FIR-bright and FIR-faint quasars, with the former showing a multiplicity rate of the order of 10 per cent (B. Trakhtenbrot, private communication), well below the 30 per cent reported here.

Regardless of the flux distribution, the multiplicity rates and the separations between the counterparts, both \cite{bussmann15} and \cite{hodge13} argue against chance associations and favour, instead, physically associated blends. On the other hand, based on simulated observations of SMGs, \cite{cowley15} suggest that 90 per cent of the total 850 \micron \, flux of a 5 mJy galaxy is the sum of three to six {\it physically unassociated} sources, with the multiplicity decreasing slowly as source flux increases. 

Sub-mm blank field number counts allow for an evaluation of whether the multiple sources within the SPIRE beam observed in a fraction of our fields are possible physical associations. The density of SCUBA sources at 850 \micron \, is reported in Tab. 16 of \cite{scott06}, according to which the density of sub-mm sources with fluxes above 2 mJy translates into 0.115$^{+ 0.016}_{-0.023}$ sources in a field the size of the SPIRE beam. This number drops quickly to 0.035$^{+0.01}_{-0.018}$ for S$_{850} > 3$ mJy, and to 0.009$^{+0.003}_{-0.006}$ for S$_{850} > 4$ mJy. Number counts derived from LABOCA and ALMA surveys \citep{karim13} with a higher resolution compared to SCUBA, suggest that only 0.0049$^{+0.008}_{-0.009}$ sources are expected at S$_{870} > 4.8$ mJy. The low density of sub-mm blank field counts is, therefore, in favour of physical associations between the quasar and the secondary ACA counterparts with fluxes distributions shown in red in Fig. \ref{fig:s870histo}, even for the faintest among the  ACA sources.

What is needed in order to solve the ambiguity regarding the nature of the associations is, of course, spectroscopy of the various counterparts, something that was only recently done by \cite{hayward18} for ten SMGs. Given the small number statistics it is difficult to draw firm conclusions, however their findings seem to suggest that sources are chance associations when their separation is $>$8\pp (six out of ten), while closer separations most likely indicate physical associations with $\Delta z < 0.004$ (three out of ten). Although it is difficult to directly compare these results with the ones presented here, we would like to point out that out of the nine sources with multiple counterparts in our sample, only two have counterparts further away than 8\pp \, (but within 11\pp), both of which occur for the third source of triple systems (J003011 and 234812, see Tab. \ref{tab:sample}).

\subsection{Major mergers as triggers of accretion onto SMBHs}

Studies of dust-reddened quasars at $z \sim 2$ with the HST have shown a large fraction of these systems to have actively merging hosts, with the fraction reaching or even exceeding 80 per cent \citep{urrutia08,glikman15}. However, HST studies of non-reddened $z \sim 2$ quasars, a population very similar to our own sample, find a 40 per cent incidence of distortion in their hosts, used as an indicator of ongoing merger processes), more in line with our findings, and similar to those in massive galaxies at the same redshift \citep{mechtley16,farrah17}.
 
At redshifts between 2 and 4 (the redshift limits of our sample) the angular scale ranges between 8.5 kpc/\pp \, and 7.1 kpc/\pp, meaning that the synthesised beam of $\sim$4\ppdot3 $\times$ 3\ppdot0 encompasses a region of about 35$\pm$3 kpc $\times$ 23$\pm$2 kpc around each quasar. The comparison to the \cite{bussmann15} data and the simulations discussed in Sec. \ref{sec:multi}, however, puts a more stringent upper limit on the physical separation among possible multiple counterparts in seemingly single sources, at or below a couple of arcseconds, in other words below a few to a couple of tens of kpc. Systems like those seen in \cite{bussmann15} must already be undergoing merging beyond a first pericentric passage, as the velocity of the galaxies attains its maximum the moment of the first close pass and the probability, therefore, of observing them at this precise moment is lower. Based on the above, we can in principle exclude the hypothesis that these systems are, in their majority, triggered by early stage mergers. 

Other than the clear cases of multiplicity, for which, however, a confirmation on physical association is needed, an indication of some objects undergoing merger also comes from the offsets of the ACA detections with respect to the SDSS coordinates. As already mentioned, for four of the objects with single ACA counterparts (as well as two with multiple counterparts), the sub-mm emission is centred more than 1\pp \, away from the SDSS centroid and nearer to the SPIRE coordinates, with mean offsets for those six objects of 3\ppdot59 and 1\ppdot62 between the ACA and the SDSS and SPIRE coordinates, respectively. This suggests that the sub-mm emission probably emerges from a source other than the quasar itself, quite possibly at the redshift of the quasar. If this were the case, the incidence of (late-type, as suggested by the separation) major mergers among these bright quasars would be higher by about 20 per cent with respect to what the multiplicity rates would indicate. Such major merger evens, however, would trigger accretion onto the SMBH in one of the objects and extreme star formation in the other.  

The idea that the peak of global SFR in merging systems occurs a few Myrs after the first near pass, with nuclear separations ranging from 10 - 100 kpc, is supported both by simulations and observations \citep[][and references therein]{barrows17}, with a delay of $\sim100$ Myrs between the peak of the global SFR and the onset of the AGN. Simulations also predict a decline of the global SFR in late-stage mergers with nuclear separation below 10 kpc, while nuclear SFR may continue to rise during the AGN stage, possibly leading to a correlation between \Lagn \, and SFR \citep{volonteri15}. And while we can certainly not distinguish between post-coalescence mergers and genuinely isolated "secular" systems from our data, the lack of any such correlation both in the full P16 sample and the SDSS/ACA sample discussed here (see also Sec. \ref{sec:nature}) argues against these systems being late-stage mergers. 
Nevertheless, the presence or not of such a correlation will be impacted by the source of the FIR/sub-mm emission. If star formation does not occur in the quasar hosts but in an SMG at the redshift of the quasar, as discussed above, there is no reason why \Lagn \, and SFR should correlate. And regardless of multiplicity, this seems to affect about 20 per cent of the sources of our sample.

\subsection{The nature of high star formation rates in quasar hosts}\label{sec:nature}

If the 870 \micron \, emission indeed comes from the cold dust in the host of at least a large fraction of optically-bright quasars, as the present study indicates, the measured fluxes correspond to SFRs of $>$1000 \Msun yr$^{-1}$. But can such SFRs be the result of secular processes? At $z$ 2--3, galaxies can form thick gas-rich disks \citep{genzel11} in such a way that the spheroids that result from the merging of such galaxies are rich in gas with high metallicity \citep{sommariva12,cullen14}. As a consequence, the dissipation inside the spheroids is quite high and the effective radii are small: at redshift 3, the physical sizes of galaxies are smaller than in the nearby Universe by a factor of $\sim$7 \citep{trujillo07, fathi12}. Thus, large SFRs are expected at those redshifts.

To check this idea, we run the self-consistent Analytic Model of Intergalactic-medium and GAlaxy \citep[AMIGA;][]{manrique15} which successfully recovers the observed properties of the high-$z$ universe \citep{salvador17}. The idea is not to do an in-depth study of the results of the run, as this will be part of a dedicated upcoming work, but rather to simply test whether there is any theoretical support for such a claim. The results of the run indeed confirm that, at redshifts between 1.5 and 3.5, the effects mentioned above lead indeed to SFRs well above a thousand \Msun yr$^{-1}$ in galaxies with \Lagn \, ranging between 10$^{45.6}$ and 10$^{46.9}$ erg s$^{-1}$, that coincide almost exactly with the limits of our sample.

\subsection{Concluding remarks}\label{sec:remarks}

The findings of this work give an unambiguous target for cosmological models to reproduce in terms of coeval SFR-AGN systems, while at the same time establishing a set of excellent laboratories for studying the relation between star formation and AGN activity in extreme systems via further observations. Sub-arcsec resolution observations are absolutely essential to a) investigate the nature of the seemingly point-like, isolated sources and determine whether they consist of one or more counterparts and b) check whether the companion sources identified in 30 per cent of the ACA images are at the redshifts of the quasars. Such observations are essential to establish the fraction of sources driven by secular processes (i.e. single sources), as well as to delimit the role of major mergers, be it early or late stage, as triggers of concomitant accretion onto SMBHs and intense star formation.

\section*{Acknowledgements}
This paper makes use of the following ALMA data: ADS/JAO.ALMA\#2016.2.00060.S and ADS/JAO.ALMA\#2011.1.00539.S. ALMA is a partnership of ESO (representing its member states), NSF (USA) and NINS (Japan), together with NRC (Canada), NSC and ASIAA (Taiwan), and KASI (Republic of Korea), in cooperation with the Republic of Chile. The Joint ALMA Observatory is operated by ESO, AUI/NRAO and NAOJ. This work makes use of TOPCAT, "TOPCAT \& STIL: Starlink Table/VOTable Processing Software", M. B. Taylor, in Astronomical Data Analysis Software and Systems XIV, eds. P Shopbell et al., ASP Conf. Ser. 347, p. 29, 2005. EH would like to thank A. Feltre, R. Ivison, and M. Zwaan for useful discussions, and A. Avison for his help with the ALMA simulations. Finally, we thank the anonymous referee for their comments and suggestions, the implementation of which further strengthened the results of this work.

%%%%%%%%%%%%%%%%%%%%%%%%%%%%%%%%%%%%%%%%%%%%%%%%%%


%%%%%%%%%%%%%%%%%%%%%%%%%%%%%%%%%%%%%%%%%%%%%%%%%%

%%%%%%%%%%%%%%%%% APPENDICES %%%%%%%%%%%%%%%%%%%%%

\appendix

\section{SDSS, UKIDSS/VHS, WISE and SPIRE cutouts}\label{sec:cutouts}

\begin{figure*}
\includegraphics[width = 3cm,height=3cm]{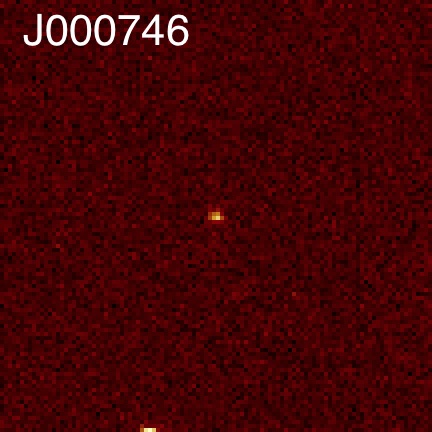}
\includegraphics[width = 3cm,height=3cm]{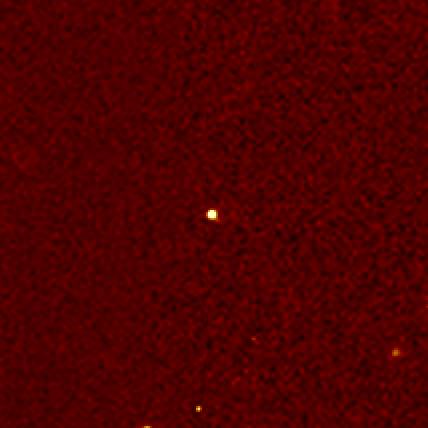}
\includegraphics[width = 3cm,height=3cm]{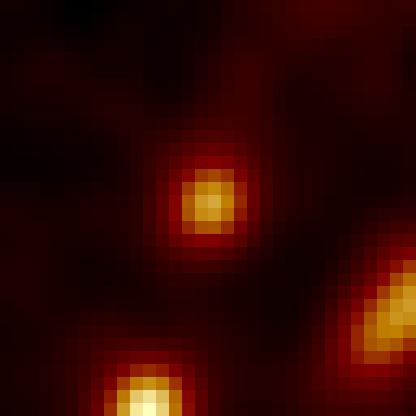}
\includegraphics[width = 3cm,height=3cm]{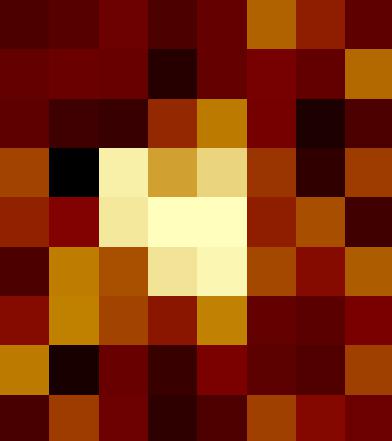}
\includegraphics[width = 3cm,height=3cm]{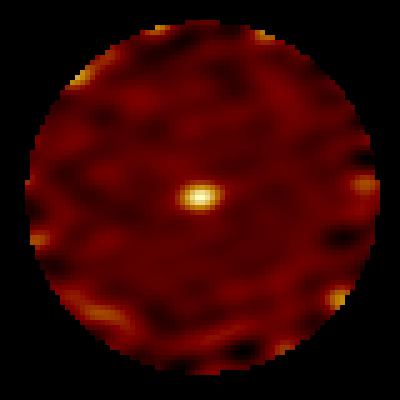} \\
\includegraphics[width = 3cm,height=3cm]{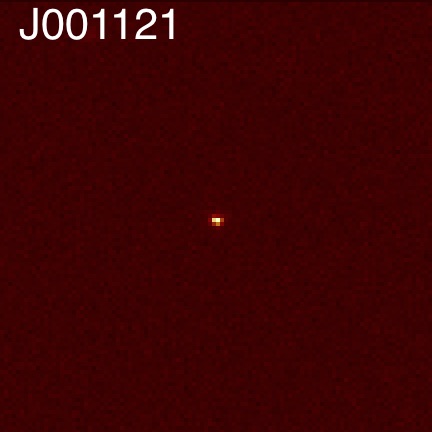}
\includegraphics[width = 3cm,height=3cm]{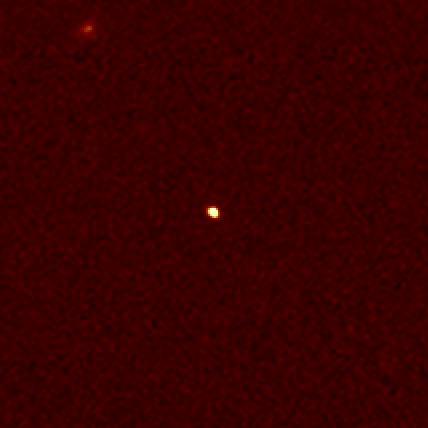}
\includegraphics[width = 3cm,height=3cm]{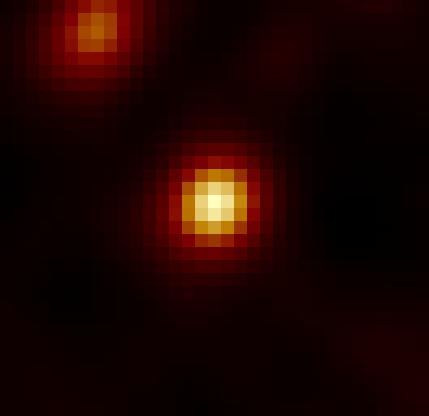}
\includegraphics[width = 3cm,height=3cm]{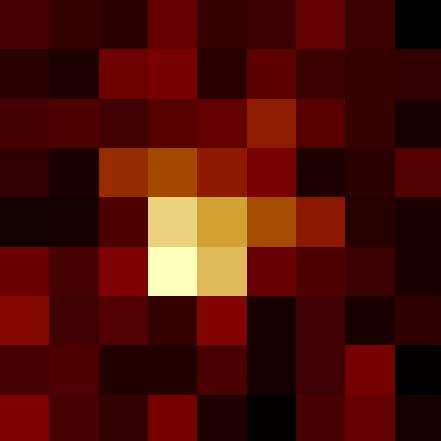}
\includegraphics[width = 3cm,height=3cm]{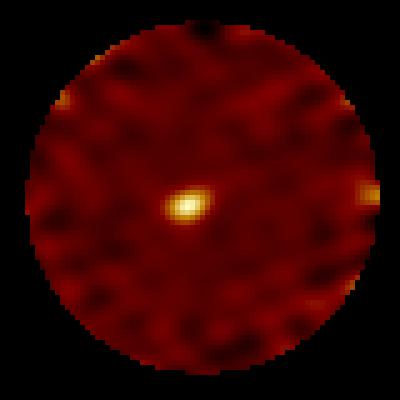} \\
\includegraphics[width = 3cm,height=3cm]{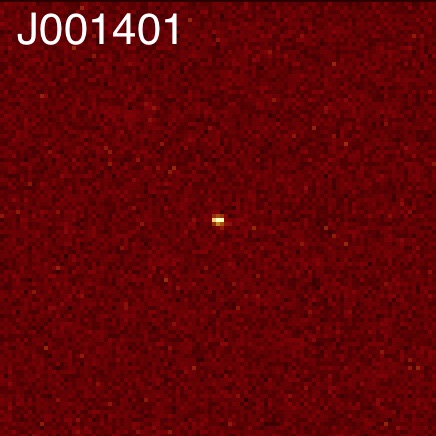}
\includegraphics[width = 3cm,height=3cm]{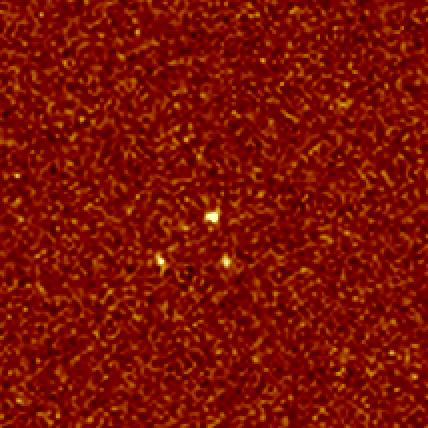}
\includegraphics[width = 3cm,height=3cm]{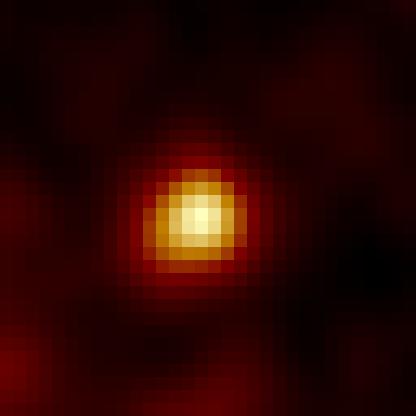}
\includegraphics[width = 3cm,height=3cm]{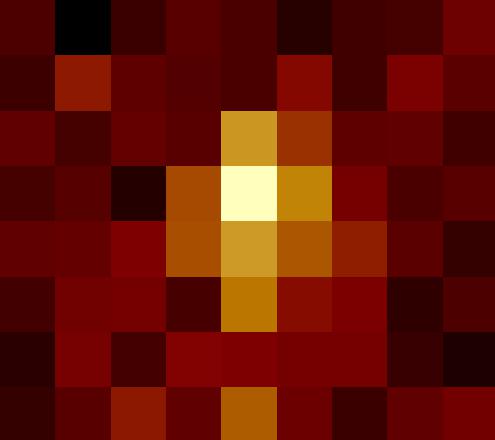}
\includegraphics[width = 3cm,height=3cm]{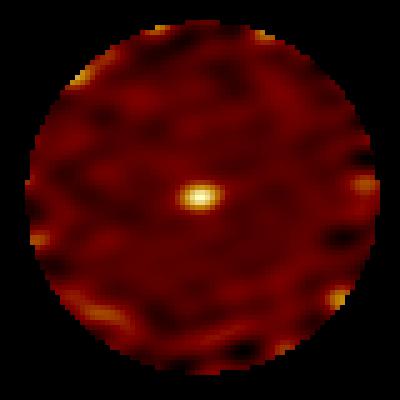} \\
\includegraphics[width = 3cm,height=3cm]{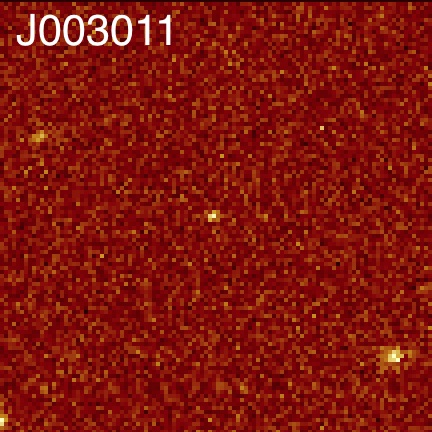}
\includegraphics[width = 3cm,height=3cm]{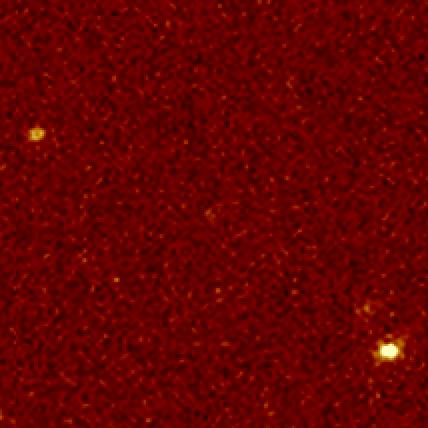}
\includegraphics[width = 3cm,height=3cm]{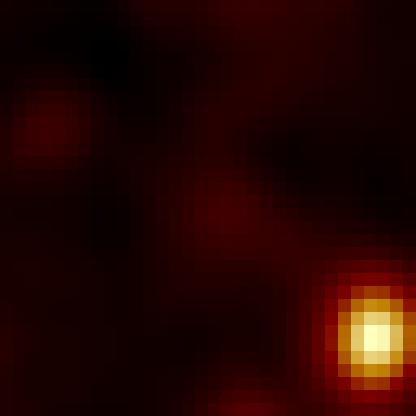}
\includegraphics[width = 3cm,height=3cm]{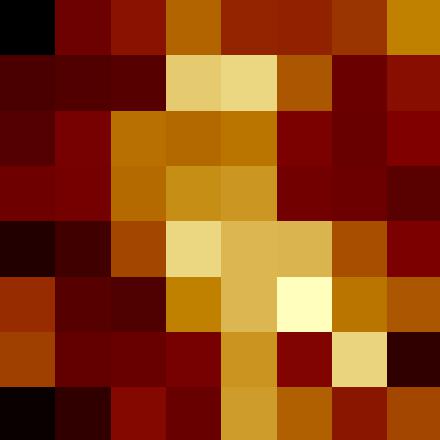}
\includegraphics[width = 3cm,height=3cm]{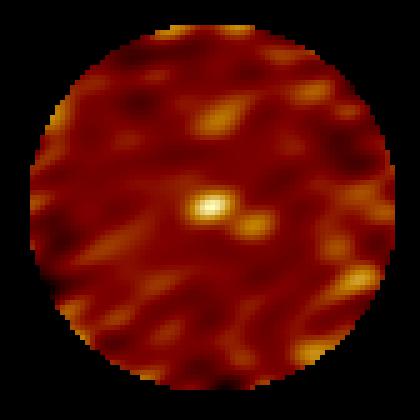} \\
\includegraphics[width = 3cm,height=3cm]{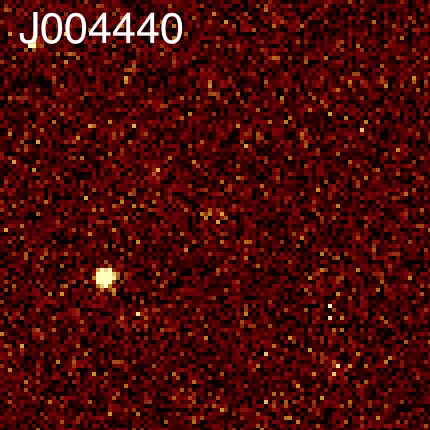}
\includegraphics[width = 3cm,height=3cm]{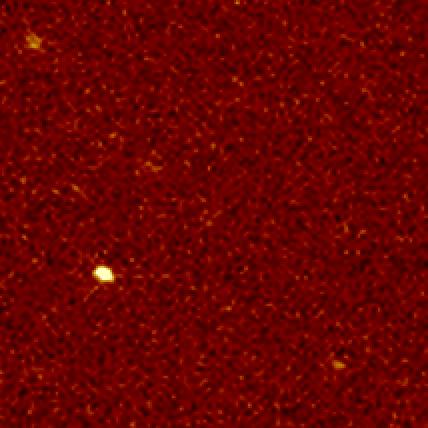}
\includegraphics[width = 3cm,height=3cm]{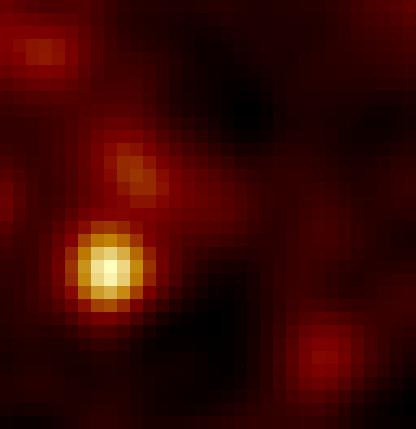}
\includegraphics[width = 3cm,height=3cm]{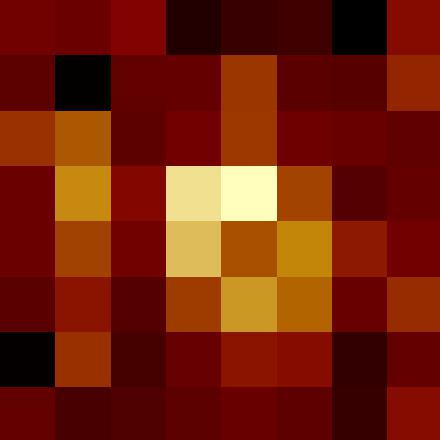}
\includegraphics[width = 3cm,height=3cm]{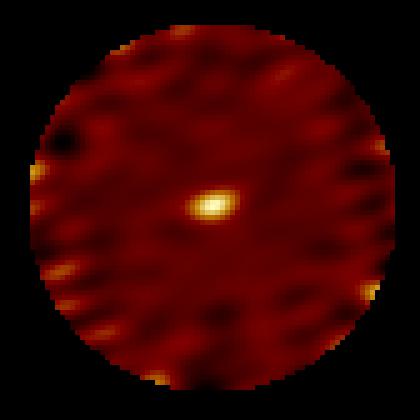} \\
\includegraphics[width = 3cm,height=3cm]{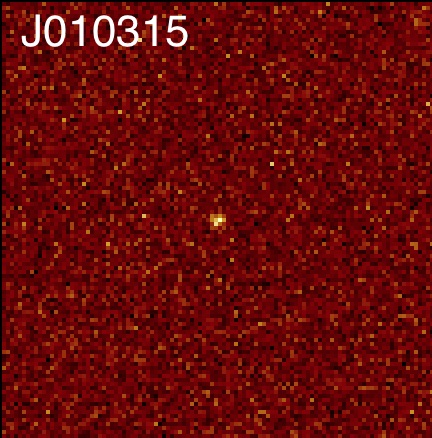}
\includegraphics[width = 3cm,height=3cm]{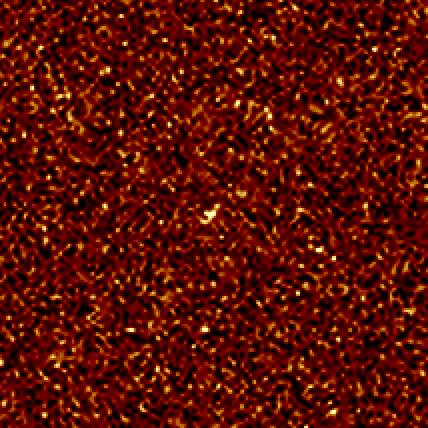}
\includegraphics[width = 3cm,height=3cm]{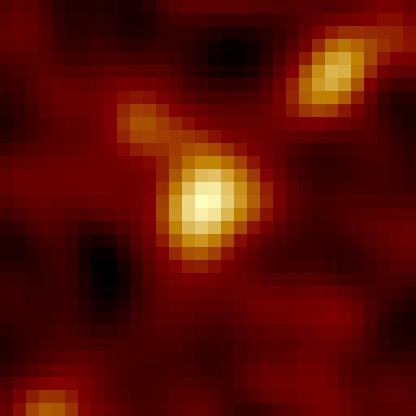}
\includegraphics[width = 3cm,height=3cm]{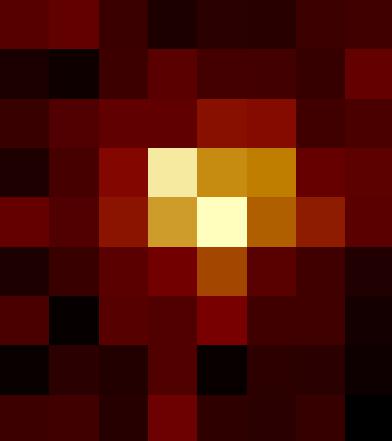}
\includegraphics[width = 3cm,height=3cm]{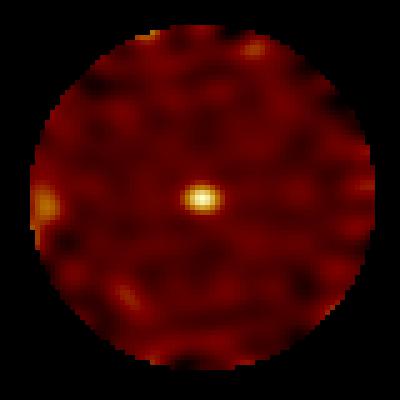} \\
\includegraphics[width = 3cm,height=3cm]{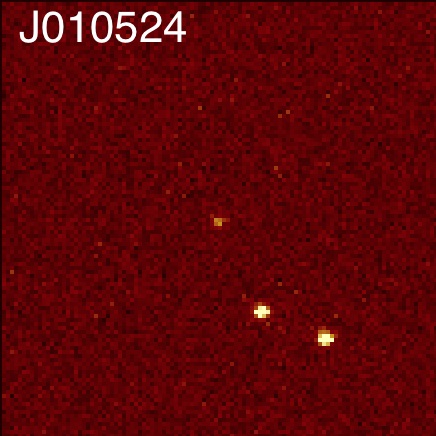}
\includegraphics[width = 3cm,height=3cm]{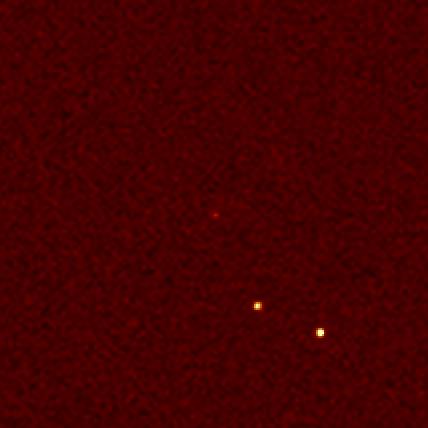}
\includegraphics[width = 3cm,height=3cm]{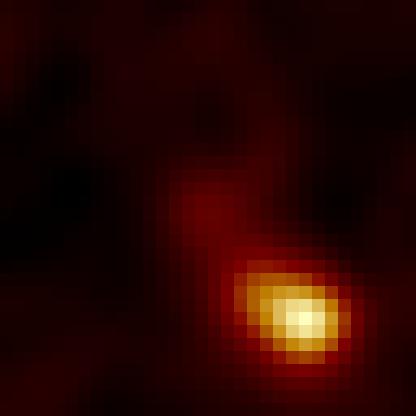}
\includegraphics[width = 3cm,height=3cm]{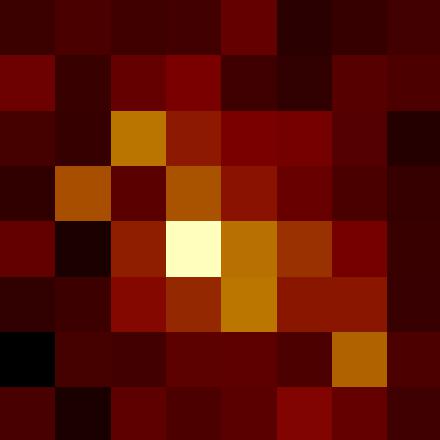}
\includegraphics[width = 3cm,height=3cm]{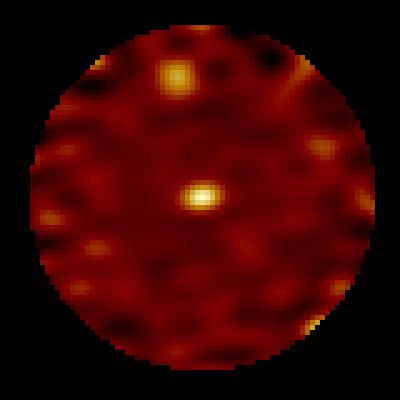}
\caption{SDSS $z$-band, UKIDSS $K$-band, WISE 3.6 \micron, SPIRE 250 \micron \, and ALMA 870 \micron \, cutouts of the same angular size (43\ppdot5 across) . For objects J33600 and J234812 VHS images are shown instead of UKIDSS. For illustration purposes, the contrast of the SDSS and NIR images has been adjusted on each individual image to enhance the features.}
\label{fig:allcutouts}
\end{figure*}

Cutouts for all the sources with the size of the ACA maps are shown in Fig. \ref{fig:allcutouts}. More specifically, from left to write we show SDSS $z$-band, UKIDSS $K_s$-band, WISE 3.6 \micron, SPIRE 250 \micron \, and the ACA maps at 870 \micron. For quasars J33600 and J234812, $K_s$ VHS cutouts are shown instead of UKIDSS as they do not fall in the UKIDSS footprint. For all cutouts, North is up and East is left.

\setcounter{figure}{0}
\renewcommand{\thefigure}{\Alph{section}\arabic{figure} (continued)}

\begin{figure*}
\includegraphics[width = 3cm,height=3cm]{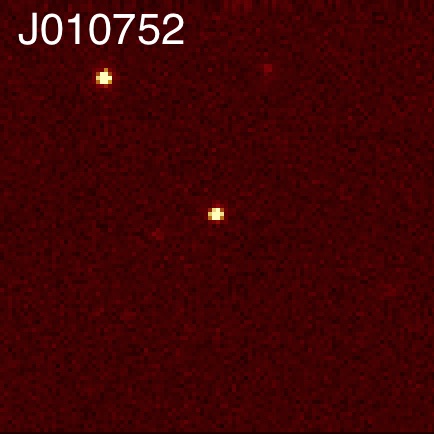}
\includegraphics[width = 3cm,height=3cm]{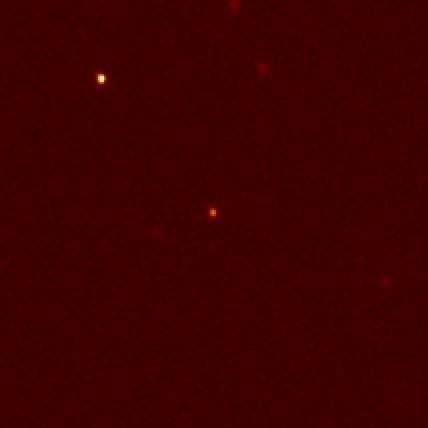}
\includegraphics[width = 3cm,height=3cm]{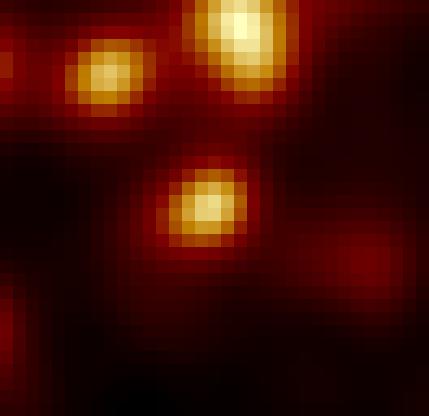}
\includegraphics[width = 3cm,height=3cm]{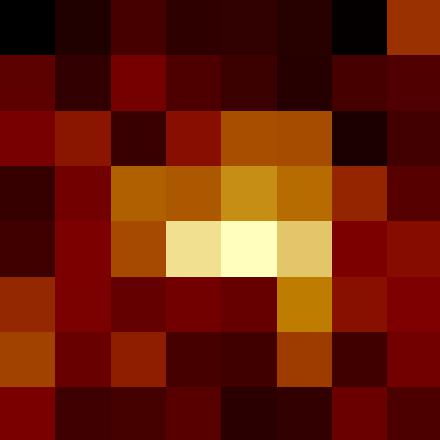}
\includegraphics[width = 3cm,height=3cm]{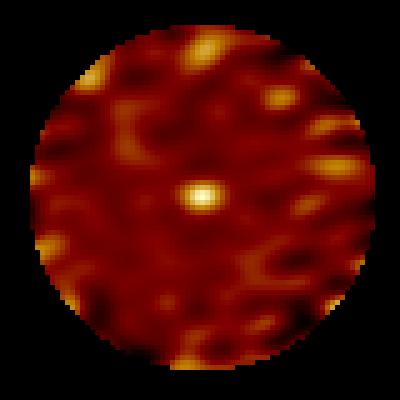} \\
\includegraphics[width = 3cm,height=3cm]{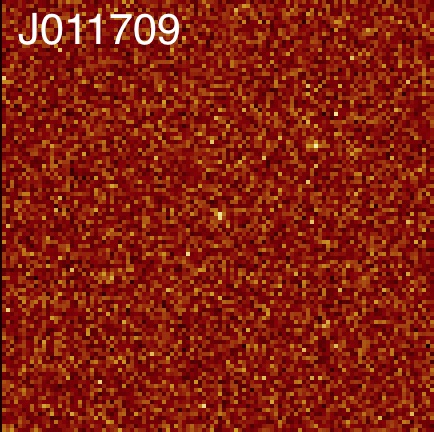}
\includegraphics[width = 3cm,height=3cm]{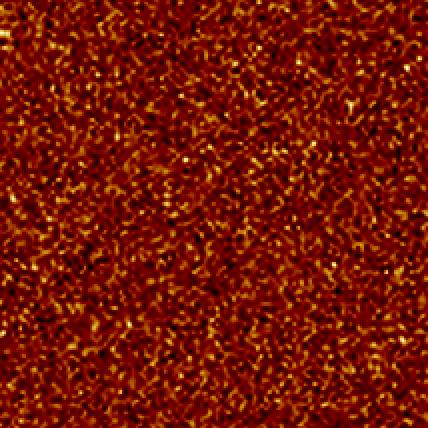}
\includegraphics[width = 3cm,height=3cm]{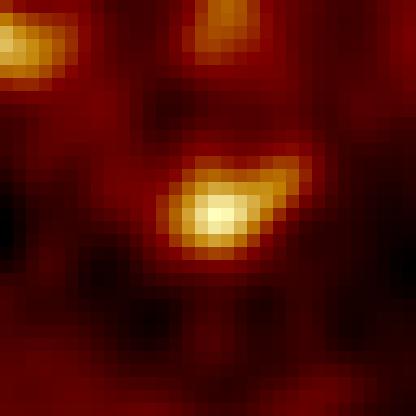}
\includegraphics[width = 3cm,height=3cm]{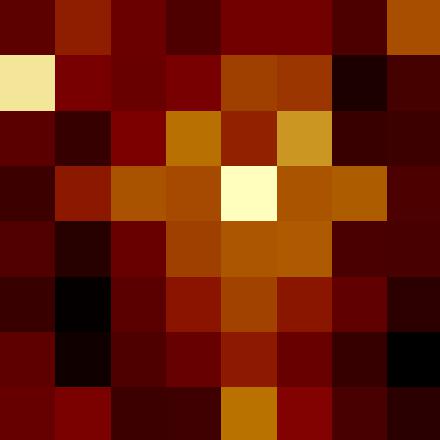}
\includegraphics[width = 3cm,height=3cm]{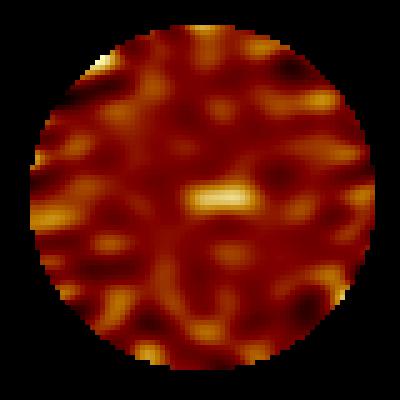} \\
\includegraphics[width = 3cm,height=3cm]{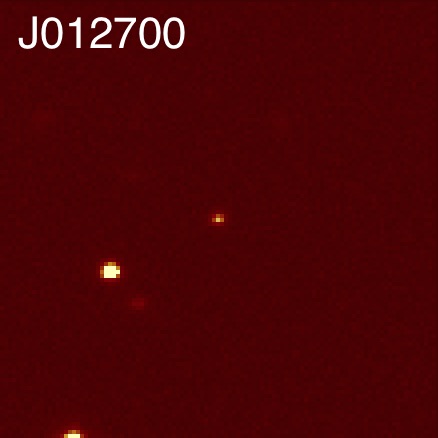}
\includegraphics[width = 3cm,height=3cm]{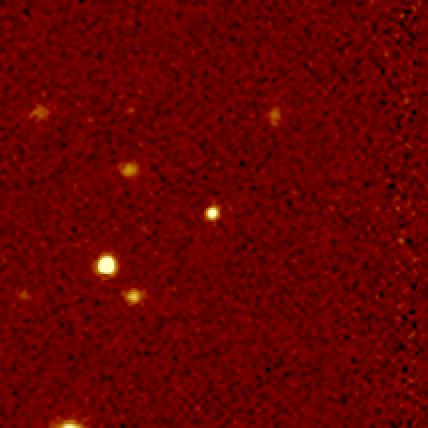}
\includegraphics[width = 3cm,height=3cm]{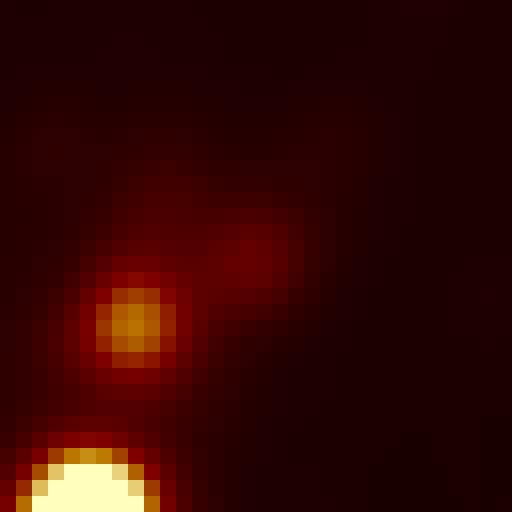}
\includegraphics[width = 3cm,height=3cm]{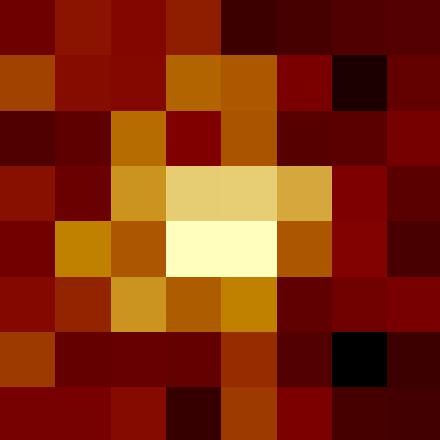}
\includegraphics[width = 3cm,height=3cm]{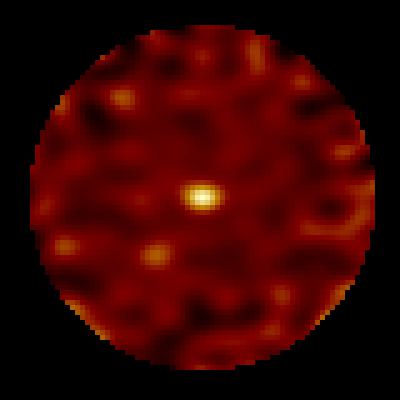} \\
\includegraphics[width = 3cm,height=3cm]{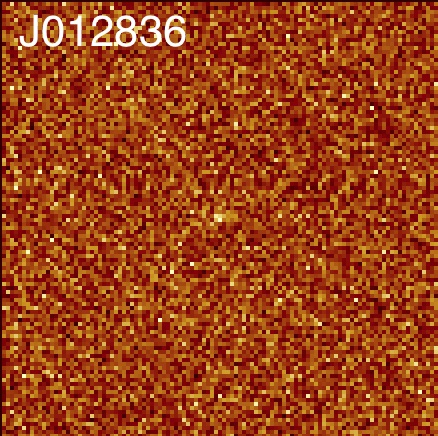}
\includegraphics[width = 3cm,height=3cm]{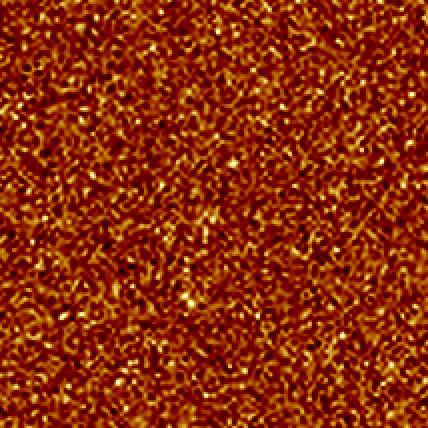}
\includegraphics[width = 3cm,height=3cm]{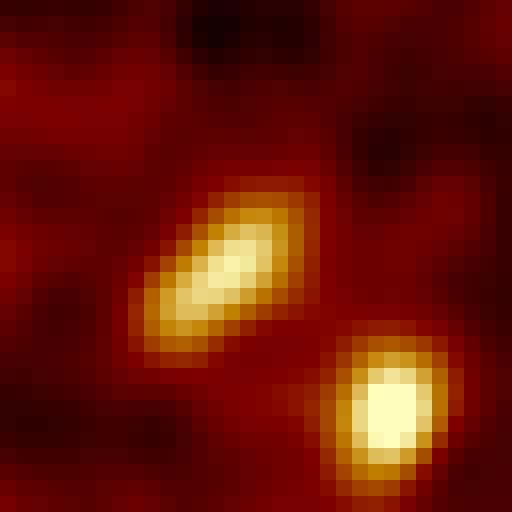}
\includegraphics[width = 3cm,height=3cm]{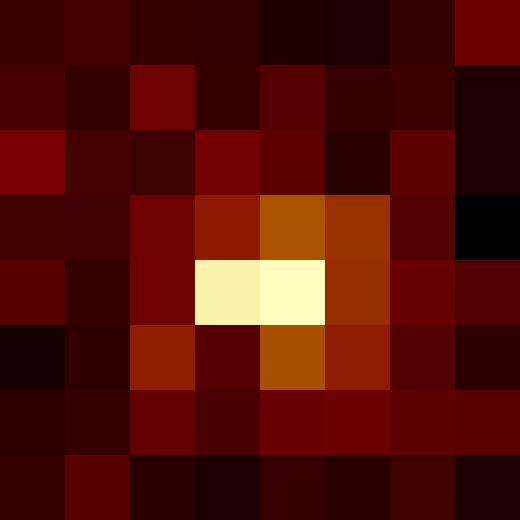}
\includegraphics[width = 3cm,height=3cm]{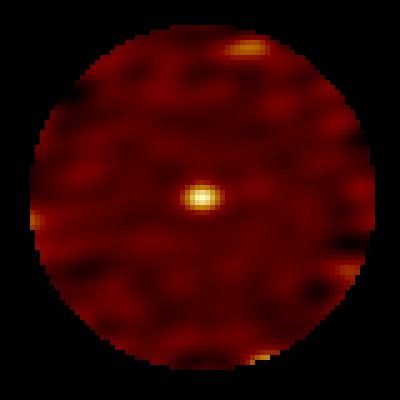} \\
\includegraphics[width = 3cm,height=3cm]{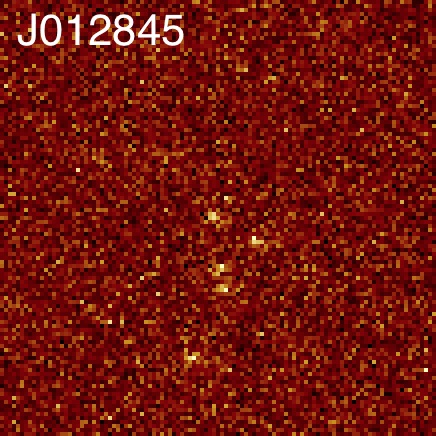}
\includegraphics[width = 3cm,height=3cm]{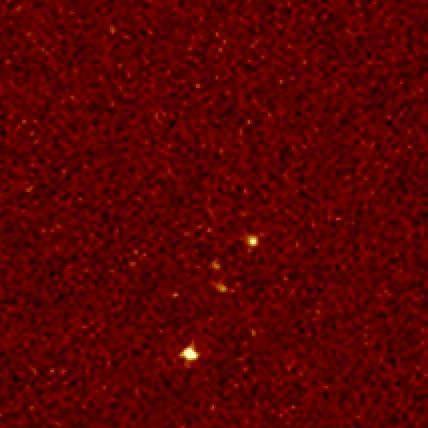}
\includegraphics[width = 3cm,height=3cm]{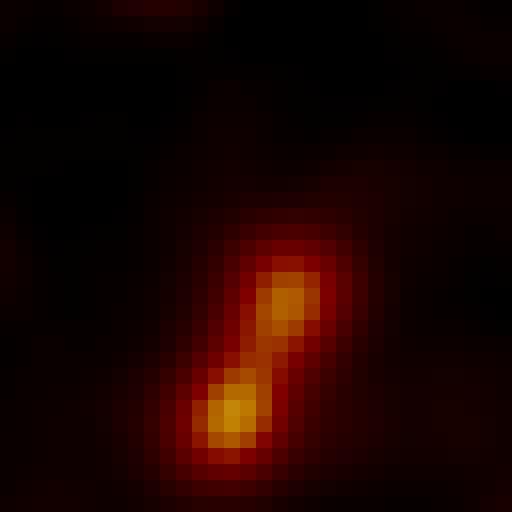}
\includegraphics[width = 3cm,height=3cm]{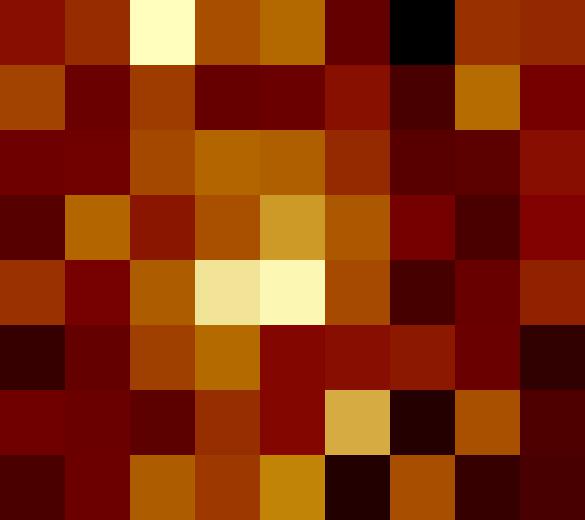}
\includegraphics[width = 3cm,height=3cm]{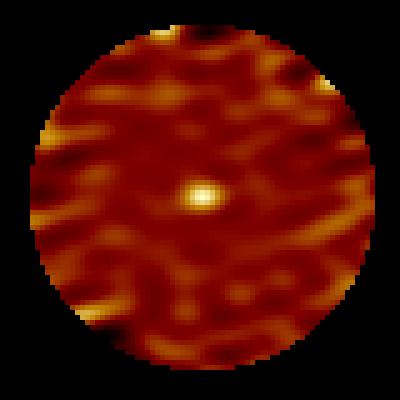} \\
\includegraphics[width = 3cm,height=3cm]{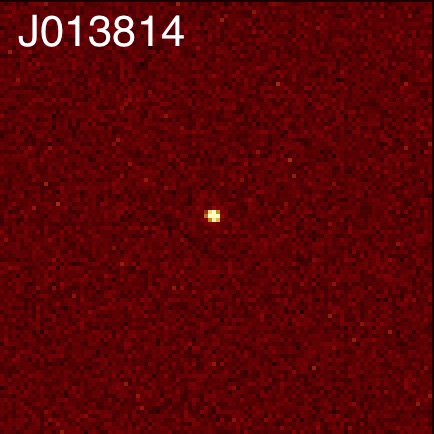}
\includegraphics[width = 3cm,height=3cm]{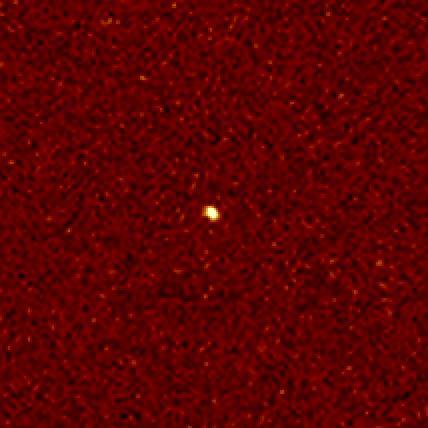}
\includegraphics[width = 3cm,height=3cm]{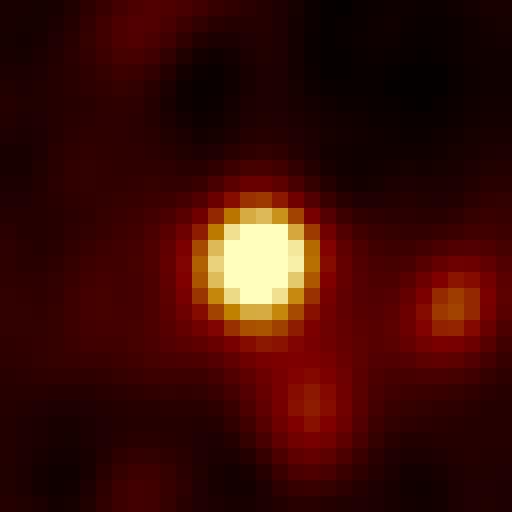}
\includegraphics[width = 3cm,height=3cm]{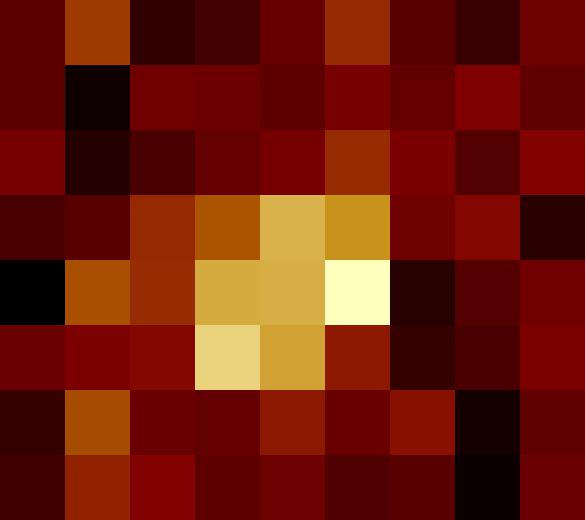}
\includegraphics[width = 3cm,height=3cm]{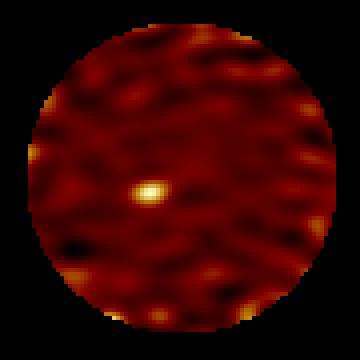} \\
\includegraphics[width = 3cm,height=3cm]{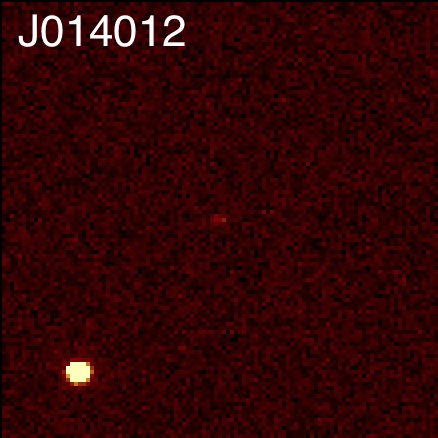}
\includegraphics[width = 3cm,height=3cm]{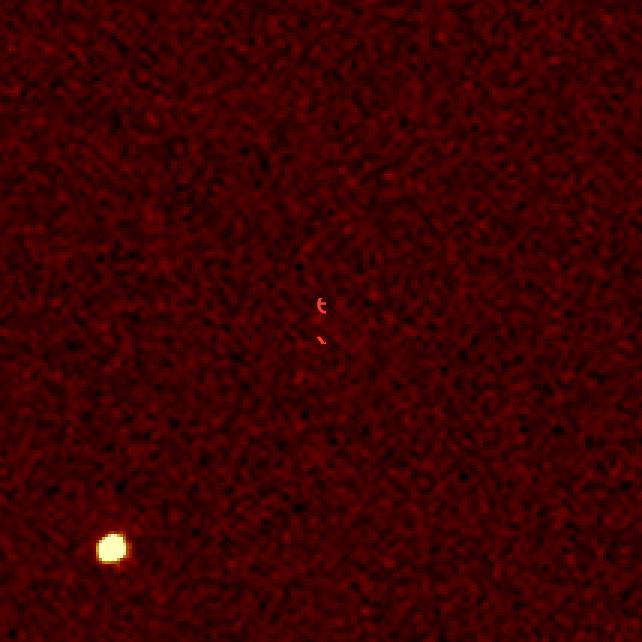}
\includegraphics[width = 3cm,height=3cm]{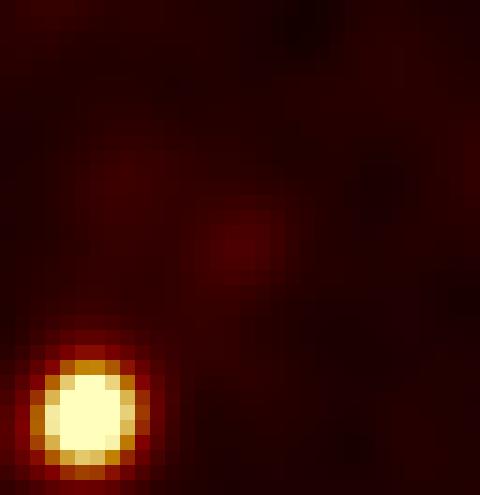}
\includegraphics[width = 3cm,height=3cm]{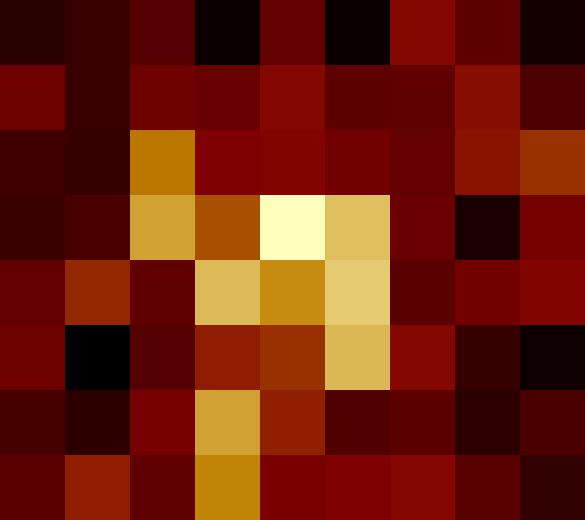}
\includegraphics[width = 3cm,height=3cm]{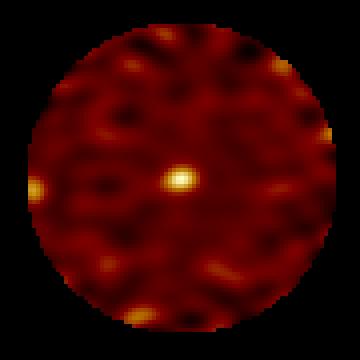}
\caption{}
\end{figure*}

\setcounter{figure}{0}
\renewcommand{\thefigure}{\Alph{section}\arabic{figure} (continued)}

\begin{figure*}
\includegraphics[width = 3cm,height=3cm]{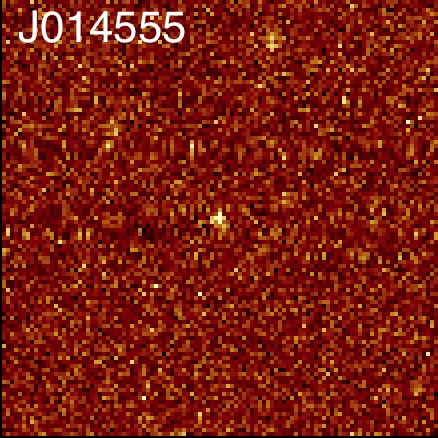}
\includegraphics[width = 3cm,height=3cm]{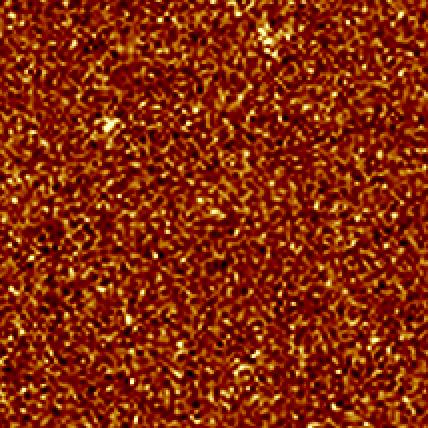}
\includegraphics[width = 3cm,height=3cm]{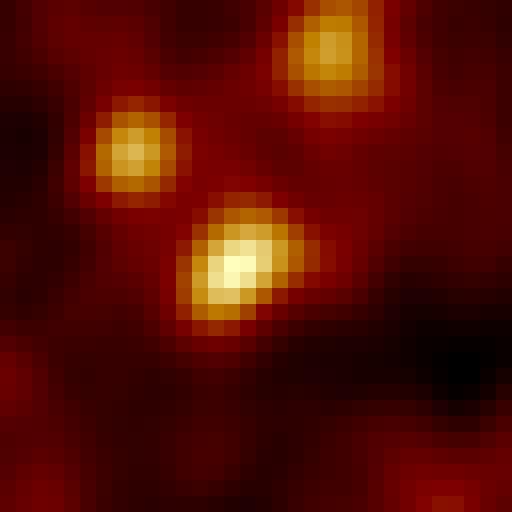}
\includegraphics[width = 3cm,height=3cm]{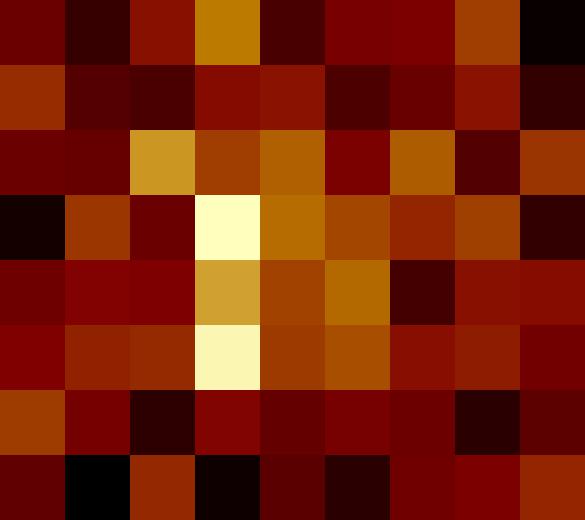}
\includegraphics[width = 3cm,height=3cm]{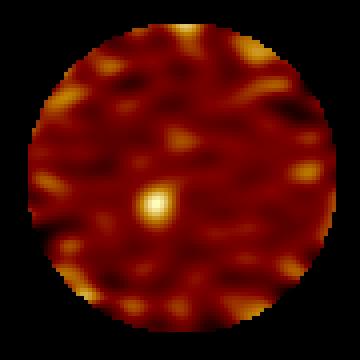} \\
\includegraphics[width = 3cm,height=3cm]{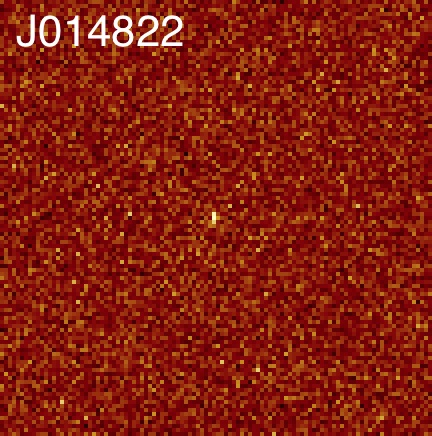}
\includegraphics[width = 3cm,height=3cm]{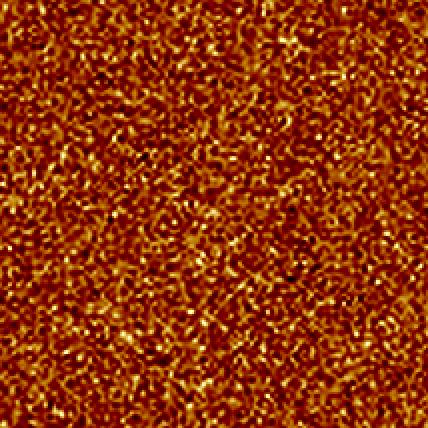}
\includegraphics[width = 3cm,height=3cm]{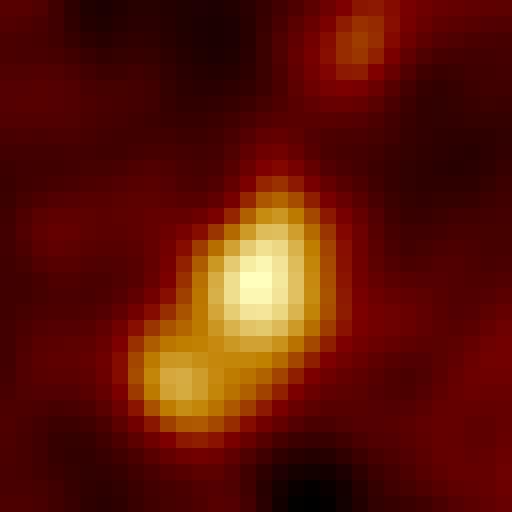}
\includegraphics[width = 3cm,height=3cm]{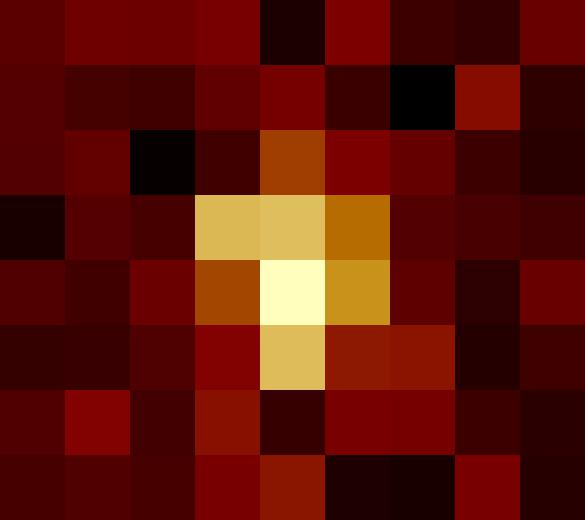}
\includegraphics[width = 3cm,height=3cm]{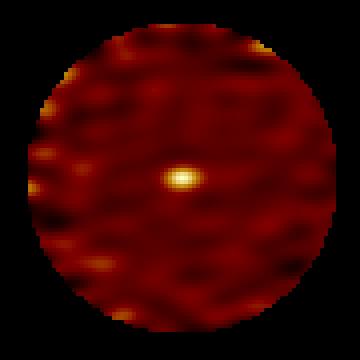} \\
\includegraphics[width = 3cm,height=3cm]{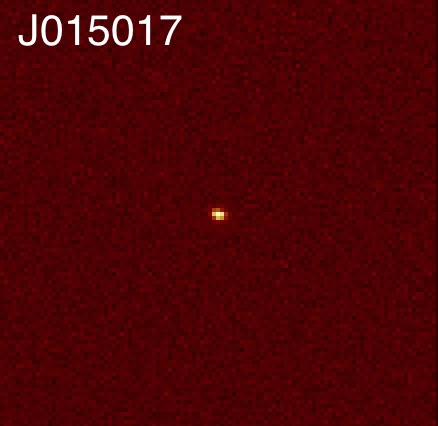}
\includegraphics[width = 3cm,height=3cm]{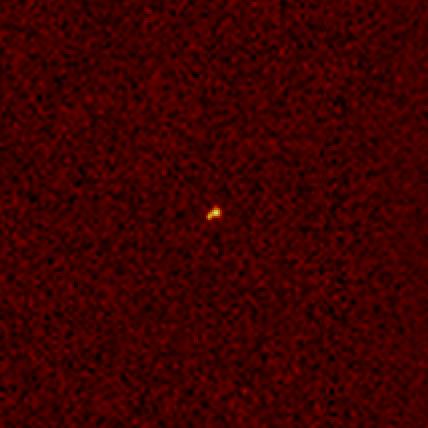}
\includegraphics[width = 3cm,height=3cm]{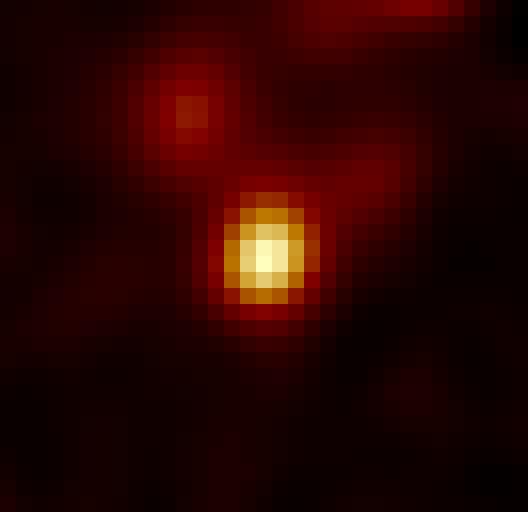}
\includegraphics[width = 3cm,height=3cm]{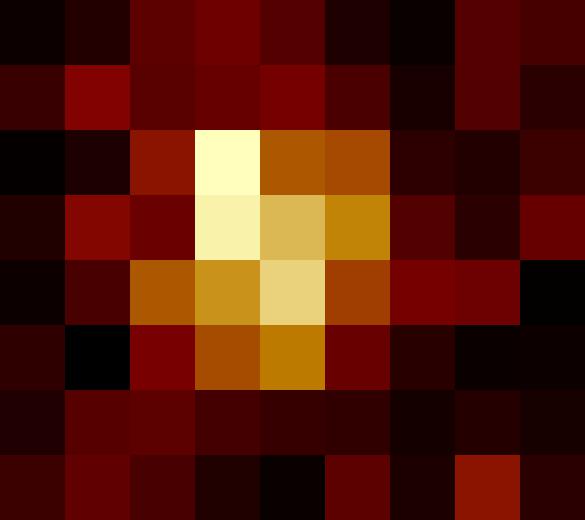}
\includegraphics[width = 3cm,height=3cm]{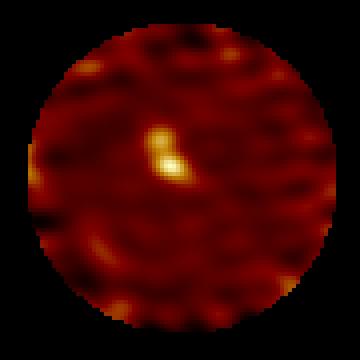} \\
\includegraphics[width = 3cm,height=3cm]{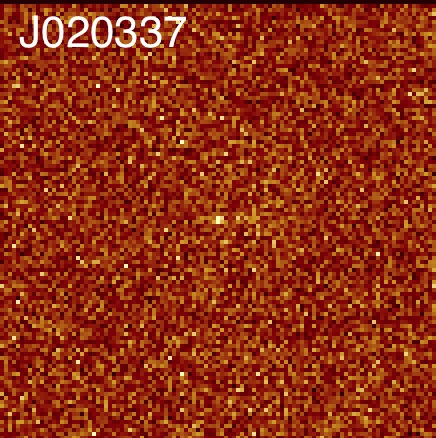}
\includegraphics[width = 3cm,height=3cm]{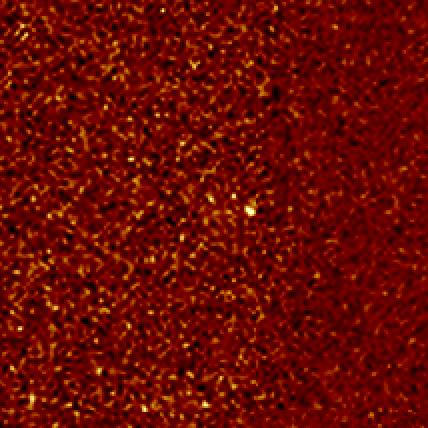}
\includegraphics[width = 3cm,height=3cm]{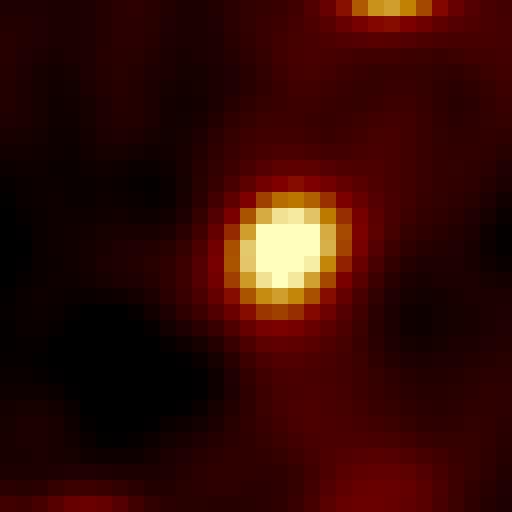}
\includegraphics[width = 3cm,height=3cm]{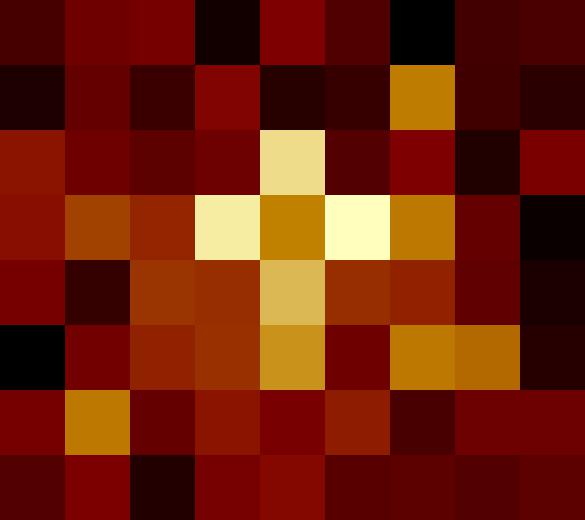}
\includegraphics[width = 3cm,height=3cm]{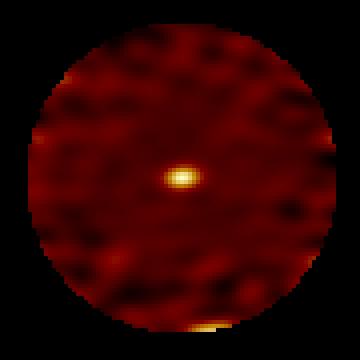} \\
\includegraphics[width = 3cm,height=3cm]{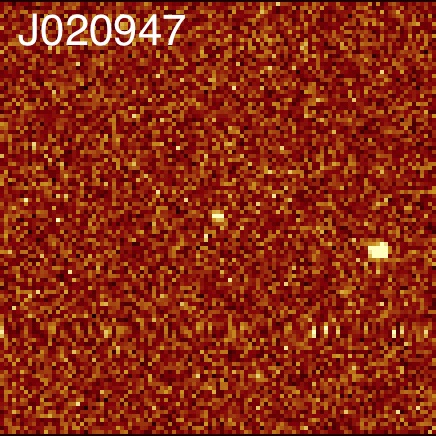}
\includegraphics[width = 3cm,height=3cm]{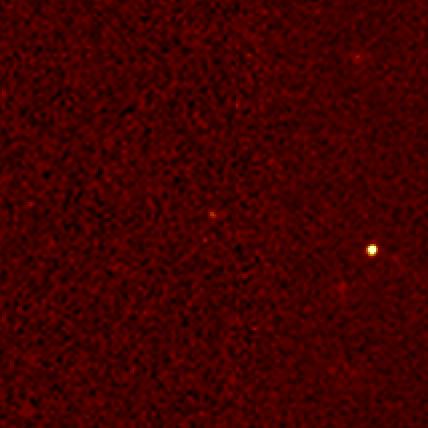}
\includegraphics[width = 3cm,height=3cm]{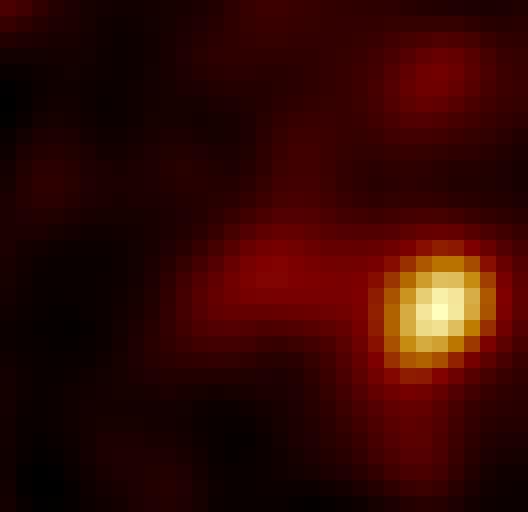}
\includegraphics[width = 3cm,height=3cm]{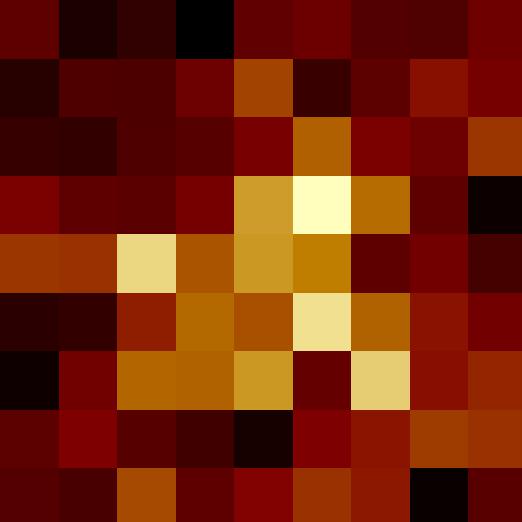}
\includegraphics[width = 3cm,height=3cm]{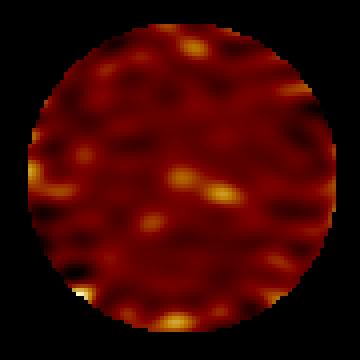} \\
\includegraphics[width = 3cm,height=3cm]{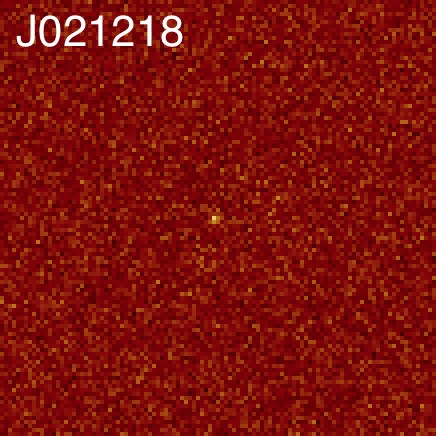}
\includegraphics[width = 3cm,height=3cm]{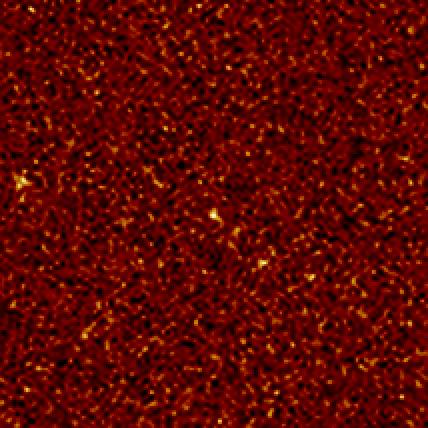}
\includegraphics[width = 3cm,height=3cm]{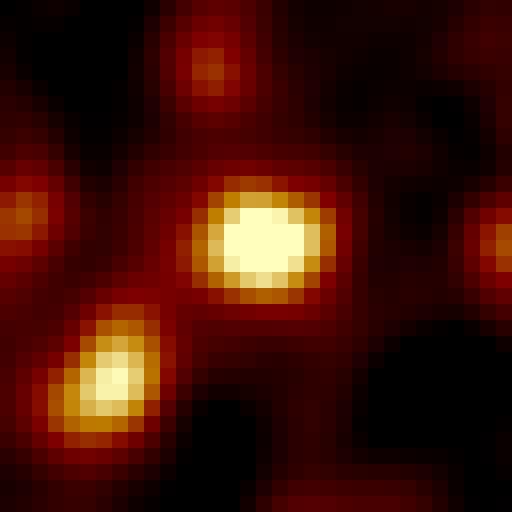}
\includegraphics[width = 3cm,height=3cm]{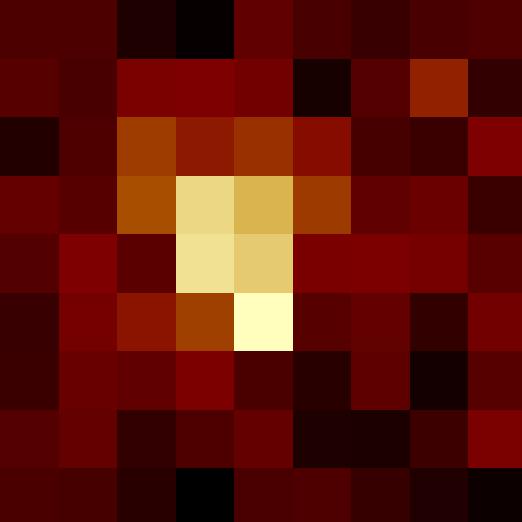}
\includegraphics[width = 3cm,height=3cm]{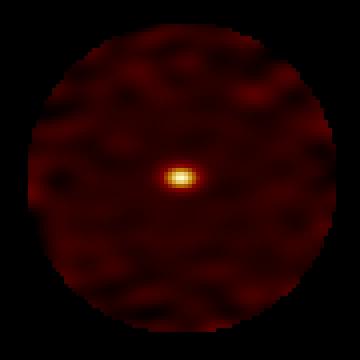} \\
\includegraphics[width = 3cm,height=3cm]{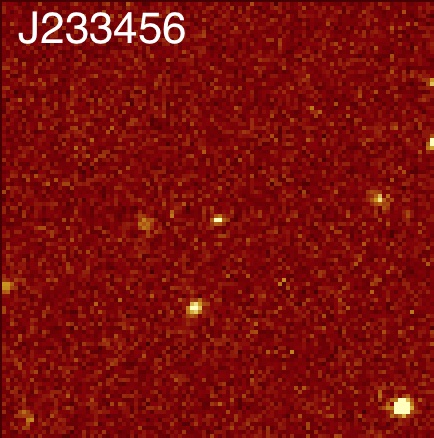}
\includegraphics[width = 3cm,height=3cm]{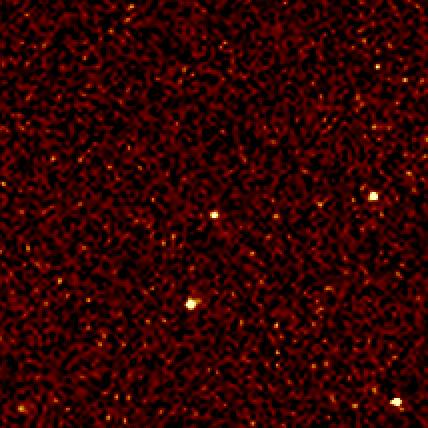}
\includegraphics[width = 3cm,height=3cm]{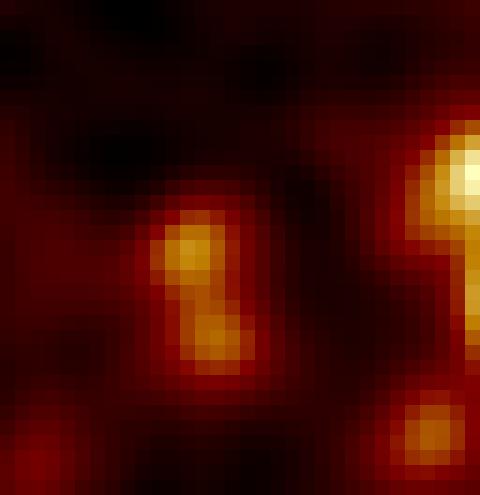}
\includegraphics[width = 3cm,height=3cm]{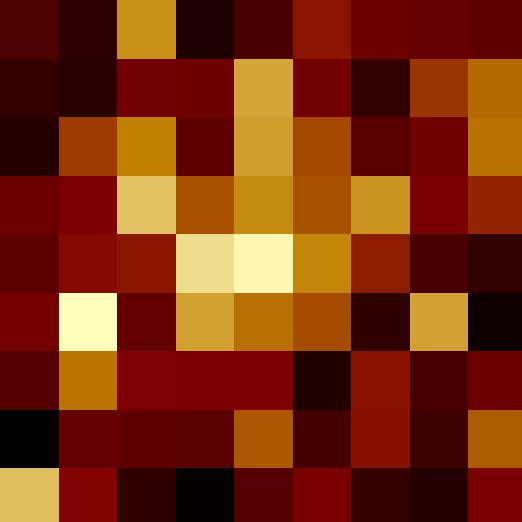}
\includegraphics[width = 3cm,height=3cm]{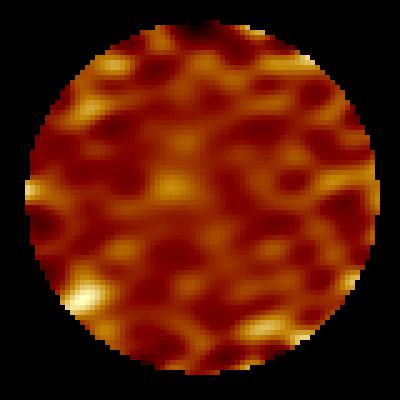}
\caption{}
\end{figure*}

\setcounter{figure}{0}
\renewcommand{\thefigure}{\Alph{section}\arabic{figure} (continued)}

\begin{figure*}
\includegraphics[width = 3cm,height=3cm]{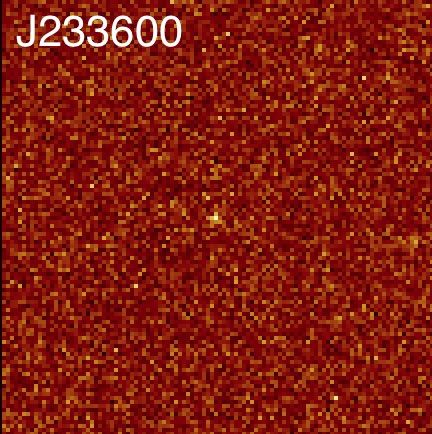}
\includegraphics[width = 3cm,height=3cm]{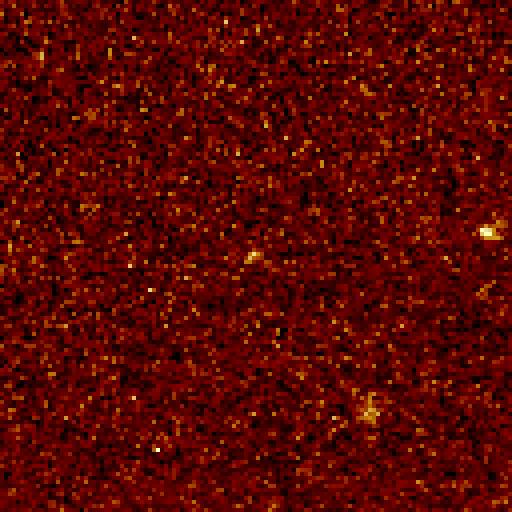}
\includegraphics[width = 3cm,height=3cm]{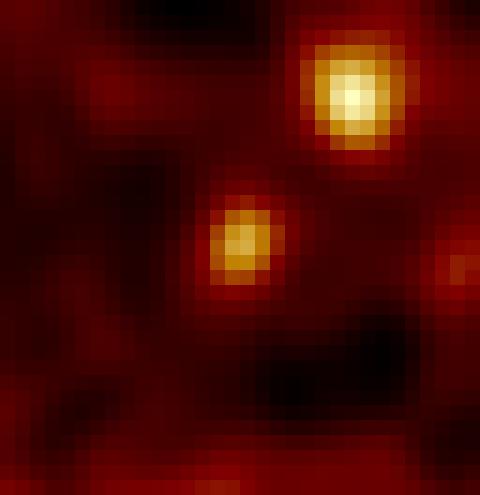}
\includegraphics[width = 3cm,height=3cm]{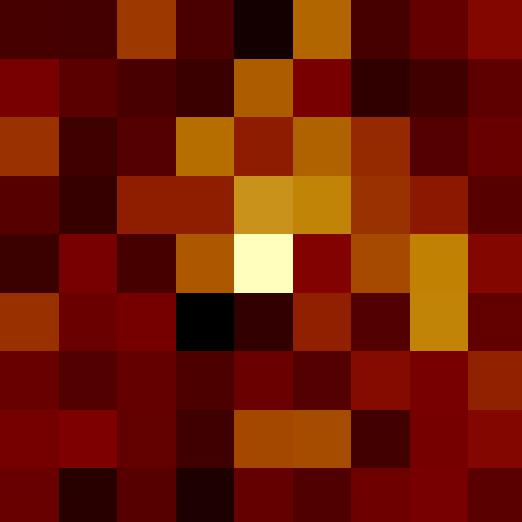}
\includegraphics[width = 3cm,height=3cm]{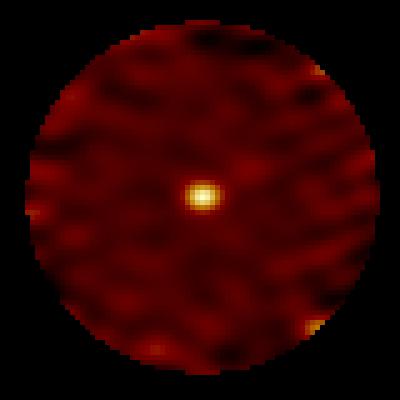} \\
\includegraphics[width = 3cm,height=3cm]{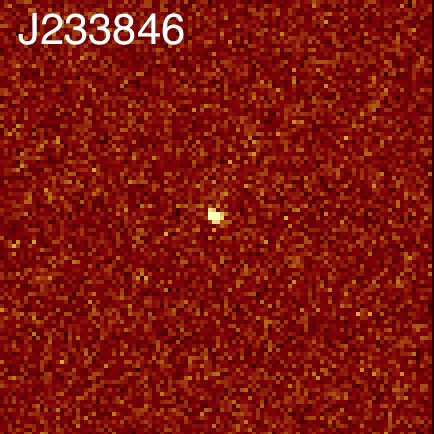}
\includegraphics[width = 3cm,height=3cm]{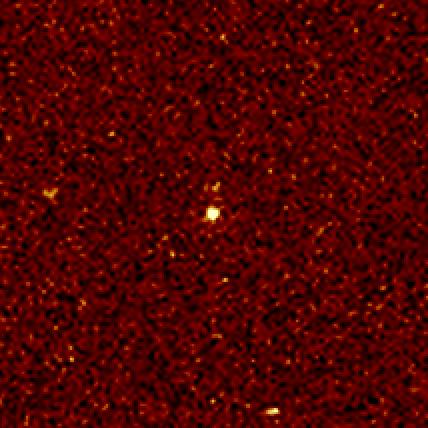}
\includegraphics[width = 3cm,height=3cm]{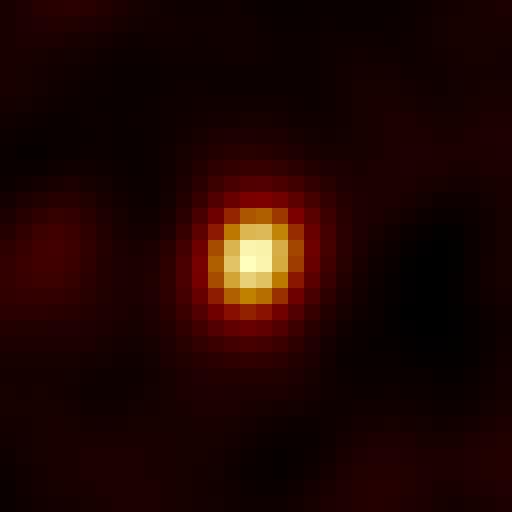}
\includegraphics[width = 3cm,height=3cm]{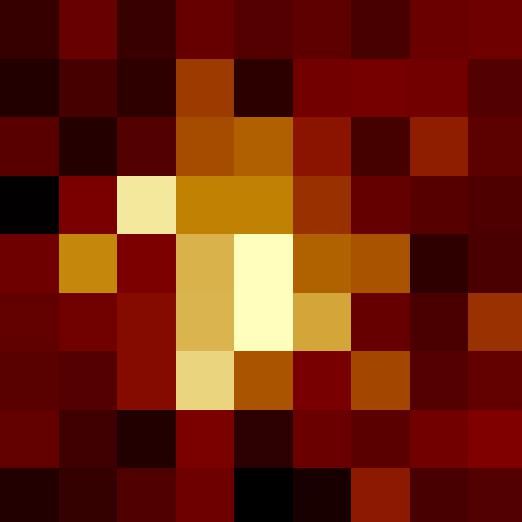}
\includegraphics[width = 3cm,height=3cm]{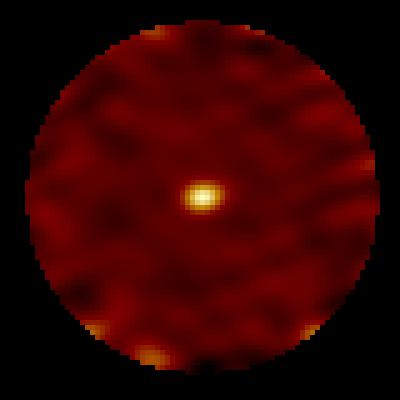} \\
\includegraphics[width = 3cm,height=3cm]{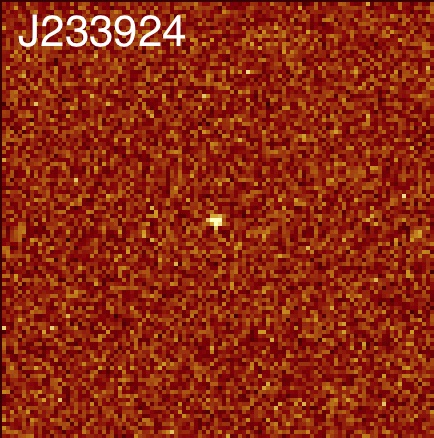}
\includegraphics[width = 3cm,height=3cm]{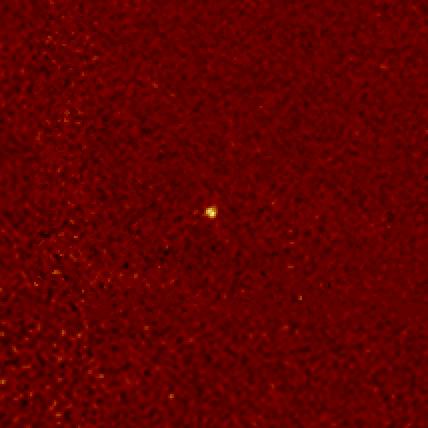}
\includegraphics[width = 3cm,height=3cm]{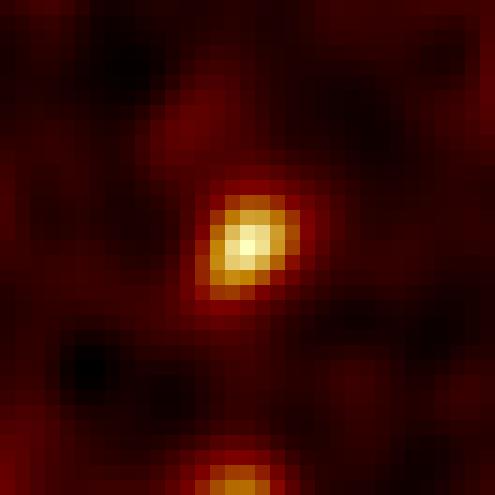}
\includegraphics[width = 3cm,height=3cm]{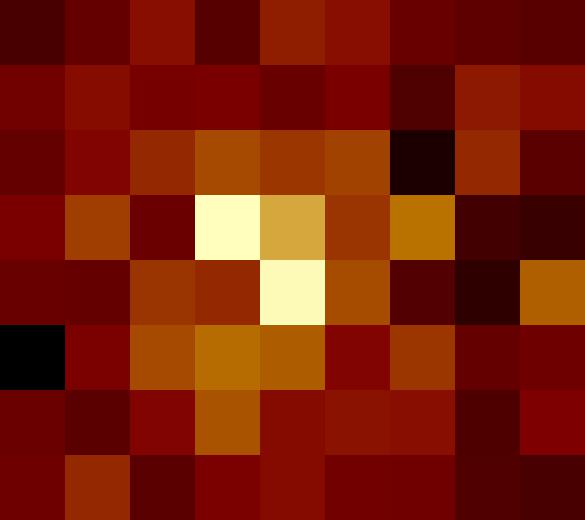}
\includegraphics[width = 3cm,height=3cm]{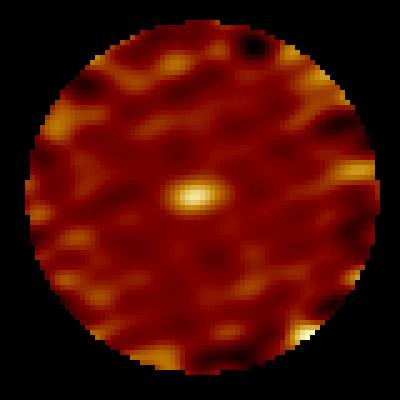} \\
\includegraphics[width = 3cm,height=3cm]{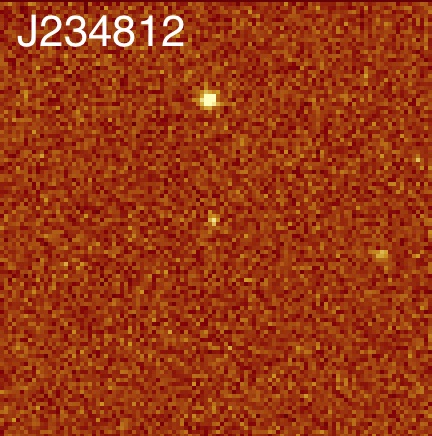}
\includegraphics[width = 3cm,height=3cm]{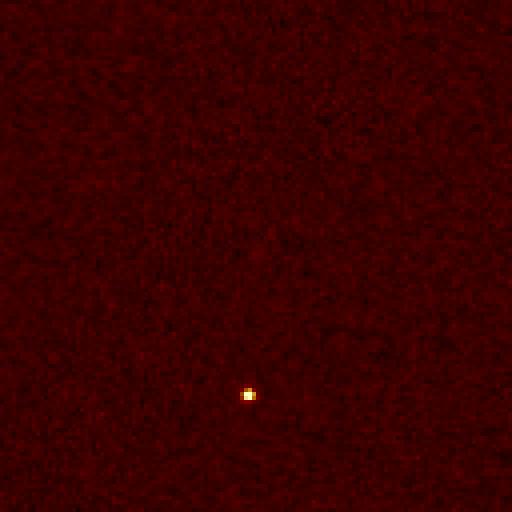}
\includegraphics[width = 3cm,height=3cm]{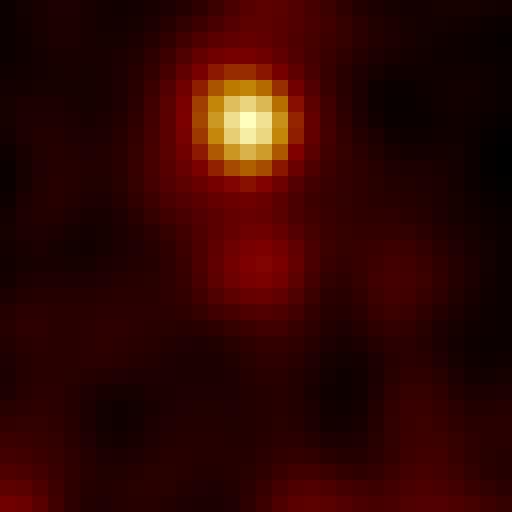}
\includegraphics[width = 3cm,height=3cm]{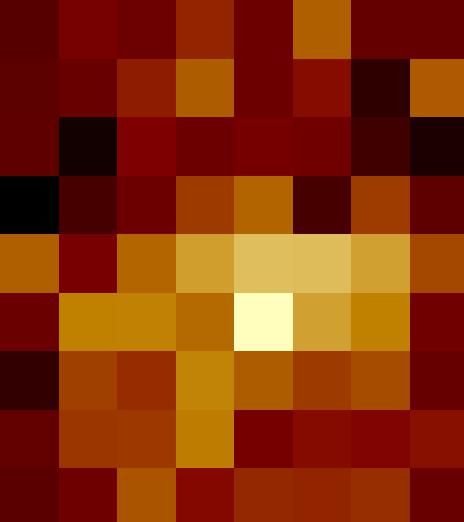}
\includegraphics[width = 3cm,height=3cm]{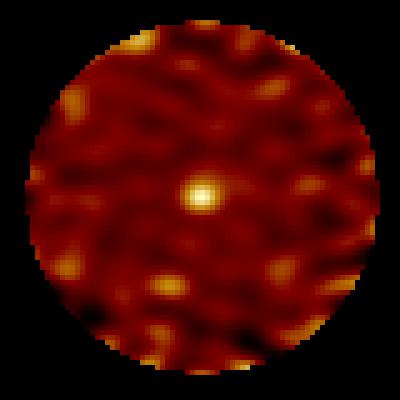} \\
\includegraphics[width = 3cm,height=3cm]{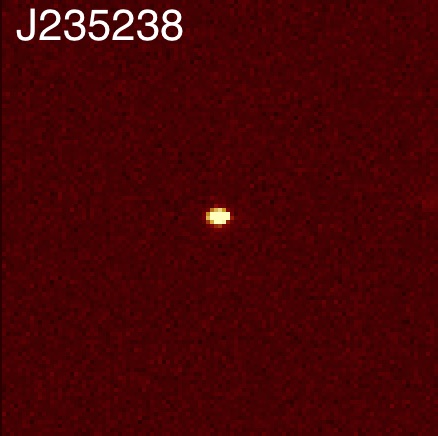}
\includegraphics[width = 3cm,height=3cm]{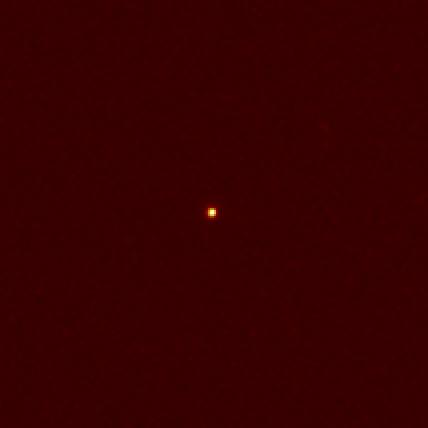}
\includegraphics[width = 3cm,height=3cm]{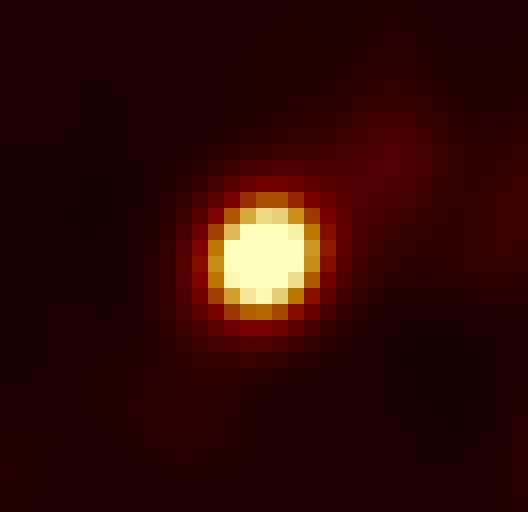}
\includegraphics[width = 3cm,height=3cm]{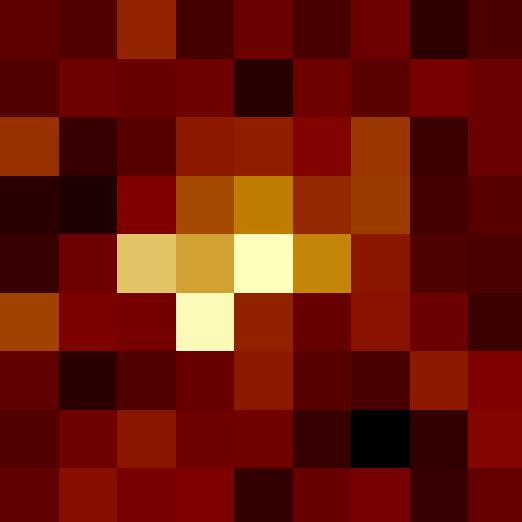}
\includegraphics[width = 3cm,height=3cm]{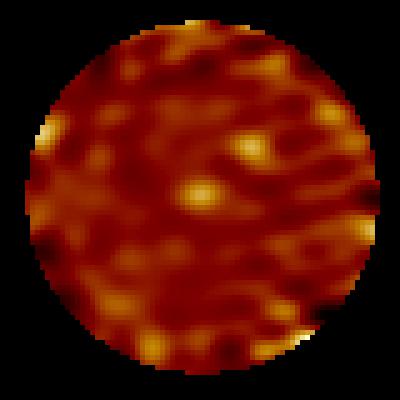} \\
\includegraphics[width = 3cm,height=3cm]{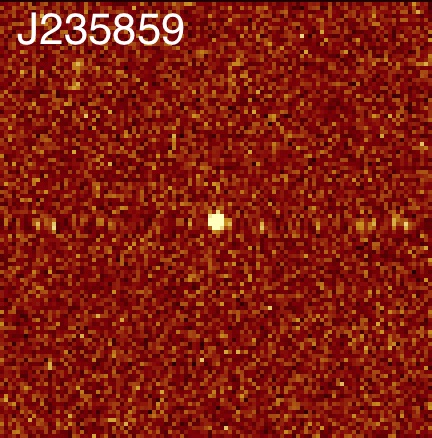}
\includegraphics[width = 3cm,height=3cm]{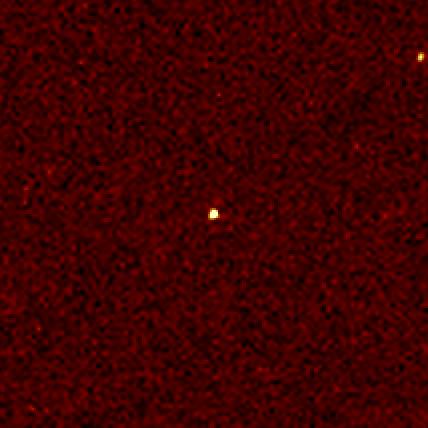}
\includegraphics[width = 3cm,height=3cm]{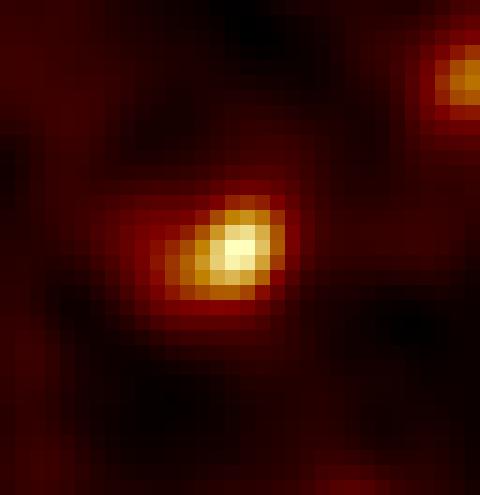}
\includegraphics[width = 3cm,height=3cm]{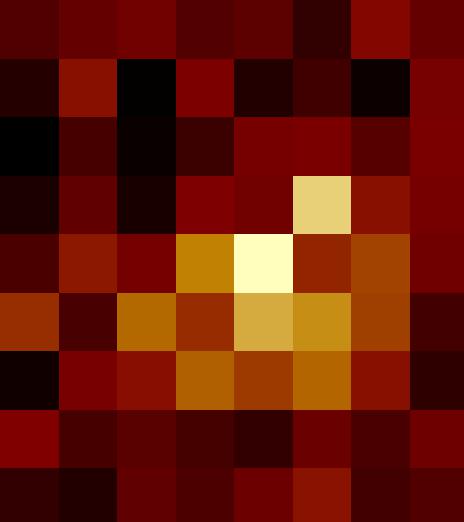}
\includegraphics[width = 3cm,height=3cm]{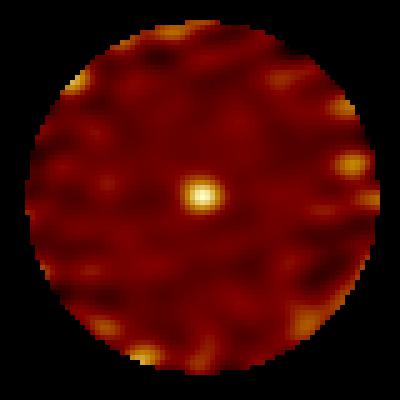} \\
\includegraphics[width = 3cm,height=3cm]{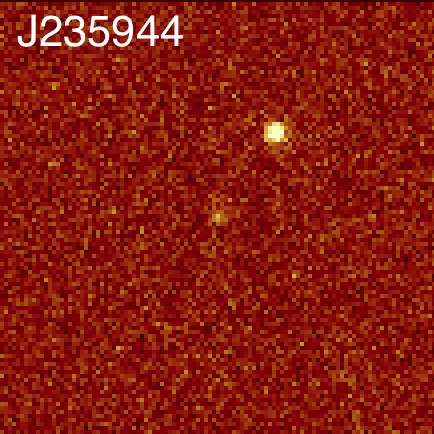}
\includegraphics[width = 3cm,height=3cm]{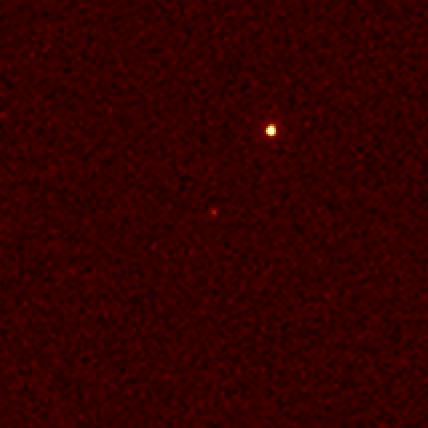}
\includegraphics[width = 3cm,height=3cm]{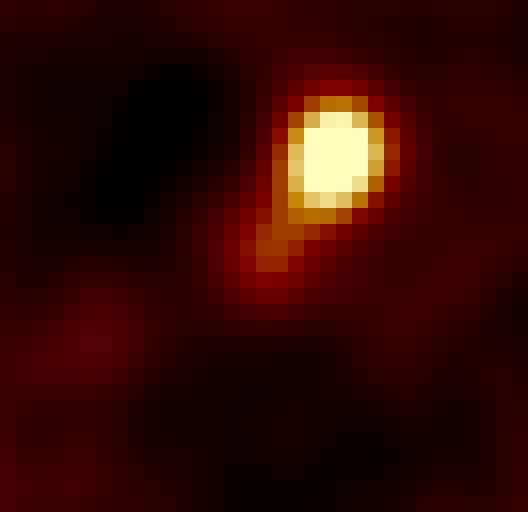}
\includegraphics[width = 3cm,height=3cm]{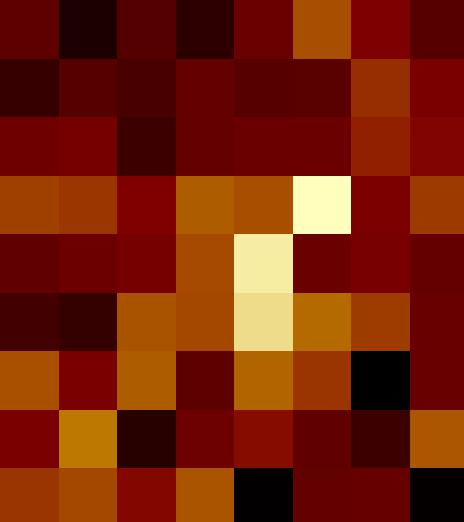}
\includegraphics[width = 3cm,height=3cm]{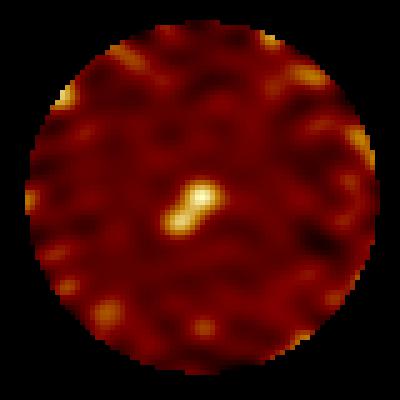}
\caption{}
\end{figure*}

\section{Other targets in the ACA maps}\label{sec:other}

In this section we present the sources in the ACA maps that lie further away than 11\pp \, from the SPIRE coordinates and are detected at above 3.5$\sigma$, that we hope will be useful for purposes like 870 \micron \, number counts. The sources are listed in Tab. \ref{tab:other}. For three of them it was not possible to extract a reliable error estimate on the flux due to their proximity to the edge of the map.
For all but one object there are no optical, near-infrared, WISE or SPIRE counterparts, as seen in the cutouts shown in Fig. \ref{fig:allcutouts} in the Appendix. The only exception is object with ACA coordinates 01:07:52.49, +01:23:55.5 that is also visible on the SDSS, UKIDSS and WISE cutouts, but there is no available information neither in SIMBAD\footnote{http://simbad.u-strasbg.fr/simbad/} nor in NED\footnote{https://ned.ipac.caltech.edu/}. The source to the east of the above detection is also visible in the optical and near-infrared images but no ACA flux was extracted as it lies at the edge of the ACA map.

\begin{table}
\begin{center}
\caption{Quasar ID in the filed of which the sources are found, coordinates and 870 \micron \, fluxes of the ACA sources that lie beyond 11$^{\prime\prime}$ from the SPIRE coordinates, detected at or above 3.5$\sigma$.}
\label{tab:other}
\begin{tabular}{lccl}
\hline
  \multicolumn{1}{c}{ID} &
  \multicolumn{2}{c}{ACA} &
  \multicolumn{1}{c}{S$_{870}$ [mJy]} \\
 & RA & Dec & \\
 \hline
J010315 & 01:03:17.02 & +00:35:23.0 & 7.50$\pm$1.10 \\
J010524 & 01:05:24.60 & --00:25:12.0 & 9.72$\pm$0.22 \\
J010752 & 01:07:52.49 & +01:23:55.5 & 5.54$\pm$0.45 \\
 & 01:07:51.39 & +01:23:40.7 & 3.64$\pm$0.45 \\
 & 01:07:51.86 & +01:23:49.5 & 2.31$\pm$0.28 \\
J011709 & 01:17:10.86 & +00:05:21.1 & 4.56 \\
J012836 & 01:28:35.98 & +00:49:52.1 & 6.66$\pm$0.63 \\
J020947 & 02:09:46.99 & +00:42:44.3 & 2.75$\pm$0.15 \\
 & 02:09:47.17 & +00:42:05.8 & 3.86 \\
J234812 & 23:48:14.09 & --03:15:10.0 & 2.41$\pm$0.41 \\
J235238 & 23:52:37.02 & +01:06:02.6 & 2.18$\pm$0.14 \\
 & 23:52:38.50 & +01:05:33.7 & 2.53 \\ 
\hline
\end{tabular}
\end{center}
\end{table}

%%%%%%%%%%%%%%%%%%%%%%%%%%%%%%%%%%%%%%%%%%%%%%%%%%

% Don't change these lines
\bsp	% typesetting comment
\label{lastpage}
\end{document}